%% file: shap0404.tex
\documentclass[10pt]{emulateapj}
\usepackage{apjfonts}
\usepackage{graphicx}
\usepackage{psfig}
\usepackage{float}

\usepackage{graphics}
\usepackage{longtable}
\usepackage{amssymb,amsmath}
\graphicspath{{Figures/}}

\newcommand{\degree}{\ensuremath{^\circ}}

\newcommand{\msun}{\ensuremath{\rm{M}_\odot}}
\newcommand{\rsun}{\ensuremath{\rm{R}_\odot}}

\newcommand{\ain}{\ensuremath{a_{\textrm{1}}}}
\newcommand{\aout}{\ensuremath{a_{\textrm{2}}}}
\newcommand{\bout}{\ensuremath{b_{\textrm{2}}}}
\newcommand{\ainf}{\ensuremath{a_{\textrm{1, f}}}}
\newcommand{\aoutf}{\ensuremath{a_{\textrm{2, f}}}}
\newcommand{\aini}{\ensuremath{a_{\textrm{1, i}}}}
\newcommand{\aouti}{\ensuremath{a_{\textrm{2, i}}}}

\newcommand{\pin}{\ensuremath{P_{\textrm{1}}}}

\newcommand{\gin}{\ensuremath{g_{\textrm{1}}}}
\newcommand{\gout}{\ensuremath{g_{\textrm{2}}}}
\newcommand{\ein}{\ensuremath{e_{\textrm{1}}}}

\newcommand{\eini}{\ensuremath{e_{\textrm{1, 0}}}}
\newcommand{\eout}{\ensuremath{e_{\textrm{2}}}}

\newcommand{\eouti}{\ensuremath{e_{\textrm{2, 0}}}}
\newcommand{\cosi}{\ensuremath{\textrm{cos} \ \textit{i}}}

\newcommand{\cosii}{\ensuremath{\textrm{cos} \ \textit{i}_{0}}}
\newcommand{\inc}{\ensuremath{i}}
\newcommand{\minprim}{\ensuremath{m_{\textrm{0}}}}
\newcommand{\minprimf}{\ensuremath{m_{\textrm{0, f}}}}
\newcommand{\minsec}{\ensuremath{m_{\textrm{1}}}}
\newcommand{\Rinprim}{\ensuremath{R_{\textrm{ms, 0}}}}
\newcommand{\Rinsec}{\ensuremath{R_{\textrm{ms, 1}}}}
\newcommand{\mout}{\ensuremath{m_{\textrm{2}}}}
\newcommand{\rperi}{\ensuremath{r_{\textrm{peri}}}}

\newcommand{\rmin}{\ensuremath{r_{\textrm{min}}}}
\newcommand{\epsoct}{\ensuremath{\epsilon_{\textrm{oct}}}}
\newcommand{\epsocti}{\ensuremath{\epsilon_{\textrm{oct, i}}}}
\newcommand{\epsoctf}{\ensuremath{\epsilon_{\textrm{oct, f}}}}

\newcommand{\tk}{\ensuremath{\textit{t}_{\textrm{K}}}}
\newcommand{\tek}{\ensuremath{\textit{t}_{\textrm{EK}}}}
\newcommand{\tgr}{\ensuremath{\textit{t}_{\textrm{GRp}}}}
\newcommand{\tms}{\ensuremath{\textit{t}_{\textrm{MS}}}}
\newcommand{\trg}{\ensuremath{\textit{t}_{\textrm{RG}}}}
\newcommand{\tafter}{\ensuremath{\textit{t}_{\textrm{after}}}}
\newcommand{\tml}{\ensuremath{\textit{t}_{\textrm{ml}}}}
\newcommand{\tmsprim}{\ensuremath{\textit{t}_{\textrm{ms, 0}}}}
\newcommand{\tmssec}{\ensuremath{\textit{t}_{\textrm{ms, 1}}}}

\newcommand{\Numgin}{\ensuremath{4}}
\newcommand{\Numein}{\ensuremath{10}}
\newcommand{\Numeout}{\ensuremath{20}}
\newcommand{\Numcosi}{\ensuremath{41}}
\newcommand{\NumTriples}{\ensuremath{\sim10^{5}}}

\newcommand{\TooCloseMS}{\ensuremath{3 \ \Rinprim}}

\newcommand{\TooCloseWD}{\ensuremath{3 \ \Rinsec}}

\newcommand{\TooCloseRG}{\ensuremath{1 \textrm{AU}}}
\newcommand{\rtide}{\ensuremath{r_{\textrm{tide}}}}

\newcommand{\EpsOctIncrease}{\ensuremath{23.3}} 
\newcommand{\EpsOctOrig}{\ensuremath{0.002}}
\newcommand{\EpsOctFinal}{\ensuremath{0.047}}
\input{Parms.tex}

\input{ParmsRtide_2.5.tex}

\input{ParmsRtide_3.5.tex}

\input{ParmsRtide_5.tex}

\input{ParmsRtide_10.tex}

\input{ParmsnoR_RG.tex}

\begin{document}

\title{The Mass-Loss Induced Eccentric Kozai Mechanism: \\A New Channel for the
Production of close compact object-stellar binaries.}
\shorttitle{The Mass-Loss Induced Eccentric Kozai Mechanism}
\shortauthors{Shappee \& Thompson}	

\author{
{Benjamin J. ~Shappee}\altaffilmark{1}
and
{Todd A.~Thompson\altaffilmark{2}}
}

\affil{Department of Astronomy, 
The Ohio State University, Columbus, Ohio 43210, USA}

\email{shappee@astronomy.ohio-state.edu, thompson@astronomy.ohio-state.edu}

\altaffiltext{1}{NSF Graduate Fellow}
\altaffiltext{2}{Center for Cosmology \& Astro-Particle Physics,
The Ohio State University, Columbus, Ohio 43210, USA}


\begin{abstract}

Over a broad range of initial inclinations and eccentricities an appreciable fraction of hierarchical triple star systems with similar masses are essentially unaffected by the Kozai-Lidov mechanism (KM) until the primary in the central binary evolves into a compact object.  Once it does, it may be much less massive than the other components in the ternary, enabling the ``eccentric Kozai mechanism (EKM):'' the mutual inclination between the inner and outer binary can flip signs driving the inner binary to very high eccentricity, leading to a close binary or collision.  We demonstrate this ``Mass-loss Induced Eccentric Kozai'' (MIEK) mechanism by considering an example system and defining an ad-hoc minimal separation between the inner two members at which tidal affects become important. For fixed initial masses and semi-major axes, but uniform distributions of eccentricity and cosine of the mutual inclination, $\sim 10 \%$ of systems interact tidally or collide while the primary is on the MS due to the KM or EKM.  Those affected by the EKM are not captured by earlier quadrupole-order secular calculations.  We show that fully $\sim30$\% of systems interact tidally or collide for the first time as the primary swells to AU scales, mostly as a result of the KM. Finally, $\sim 2 \%$ of systems interact tidally or collide for the first time after the primary sheds most of its mass and becomes a WD, mostly as a result of the MIEK mechanism. These findings motivate a more detailed study of mass-loss in triple systems and the formation of close NS/WD-MS and NS/WD-NS/WD binaries without an initial common envelope phase.

\end{abstract}


\section{Introduction}
\label{sec:introduc}

\citet{kozai62} and \citet{lidov62} showed that a hierarchical triple system
with an inner binary of masses \minprim{} and \minsec{}, and a tertiary of mass
\mout{}, can exchange angular momentum between the inner and outer orbits
periodically.  These oscillations drive the eccentricity of the inner binary
(\ein) to high values if the tertiary is inclined with $39.2 \degree \lesssim
\inc \lesssim 141.8 \degree$.  For nearly circular orbits, in the test-particle
approximation $(\minsec \ll \minprim, \mout)$, the maximum eccentricity is given
by \citep{innanen97}
\begin{equation}
 e_{\rm{in, max}} = \left( 1 - \frac{5}{3} \rm{cos}^{2}i \right)^{1/2},
 \label{eq:kozaiemax}
\end{equation}
when the three-body Hamiltonian is expanded to quadrupole order in the ratio of
the inner to outer semi-major axes ($\ain/\aout$), and neglecting tidal forces
and general relativity \citep{blaes02, miller02, fabrycky07}.  

Tidal friction in Kozai-affected triples can cause \ain{} to decrease.  As the
inner binary is driven to high eccentricity, the periastron of the secondary approaches the primary,
and tidal friction tends to circularize the orbit \citep{mazeh79}.  
This mechanism has been proposed as an explanation for the prevalence of ``hot''
Jupiters with few-day orbits around their host stars \citep{wu03, fabrycky07,
wu07, naoz11a, naoz11b}, the very high triple fraction of close solar-type
binaries with periods less than $\sim 5$ days \citep{tokovinin06, fabrycky07},
and the formation of blue stragglers in globular clusters \citep{perets09a}. 
The combination of Kozai cycles and tidal friction has also been explored
generally for triple star systems in \citet{kiseleva98} and \citet{eggleton01}.
As discussed by \citet{wu03}, for the case of Jupiter-mass planets around
Sun-like stars, by equating the tidal and Kozai forcing terms in the equation
for the secular evolution of \ein, one estimates the circularization semi-major
axis of the inner binary to be $\sim 3 \rsun$. 

\citet{lithwick11}, \citet{katz11}, and \citet{naoz11a, naoz11b} have recently
emphasized that the commonly used quadrupole-order expansion of the three-body Hamiltonian is
insufficient to capture the secular dynamics of triple systems in the test
particle approximation when the outer eccentricity (\eout{}) is non-zero. The
parameter
\begin{equation}
 \epsoct = \left( \frac{\minprim - \minsec}{\minprim + \minsec} \right) \left(
\frac{\ain}{\aout} \right) \frac{\eout}{1 - \eout^{2}}
 \label{eq:epsoct}
\end{equation}
measures the importance of the octupole-order terms in the doubly-averaged
three-body Hamiltonian relative to the quadrupole-order terms. \citet{naoz11a} first showed that when $(\minsec \ll \minprim,
\mout)$ and $\eout \neq 0$, it is possible for the triple to ``flip'': the system
exhibits quasi-periodic cycles in \cosi{} through 0, and the tertiary passes
from prograde to retrograde and vice versa.  These flips occur even for
$\epsoct{}$ as small as $10^{-3}$ in the test particle approximation, provided
that the system is sufficiently inclined initially and the arguments of periastron of the inner
orbit (\gin) and the outer orbit (\gout) are chosen judiciously (see Figures 7
and 3 of \citealp{lithwick11} and \citealp{katz11}, respectively).  During these
flips, the inner binary is driven to extremely high eccentricities ($1-\ein \sim
10^{-5}$).  This behavior occurs for a broad range of parameters in
octupole-order calculations, but it is not present in the quadrupole-order
calculations, and has been referred to as the ``eccentric Kozai mechanism.''

An important piece of physics missing from the test particle approximation of
the eccentric Kozai mechanism is the $(\minprim - \minsec)$ term in \epsoct{}
\citep{krymolowski99, ford00, blaes02, naoz11b}. Because the three-body
Hamiltonian in Jacobi coordinates, expanded to all orders in $\aout / \ain$  has
a sum over $(\minprim^{j-1} - (-\minsec)^{j-1})$ for $j=2,3,\dotsm$, when
$\minprim = \minsec$ all $j=\rm{odd}$ terms are zero \citep{harrington68}. 
Importantly, the flip phenomenon present in the test-particle calculations at
modest \eout{} and \inc{} is strongly suppressed as $\minprim \rightarrow
\minsec$
because the octupole-order terms become negligible\footnote{Intuitively, any odd-order terms in the multipole expansion of the three-body Hamiltonian are odd functions of $r_{01}$, the separation vector between members of the inner binary, and thus averages to zero for symmetric inner masses.}.
Our own experiments with the octupole-order code based on 
\citet{blaes02} and \citet{thompson11} shows that in order for flips to occur over a 
broad range in \gin{} and \gout{}, \minsec{} must be less than \minprim{} by 
a factor of $\sim 2$ for a given $\ain / \aout$ and $\eout$.  

The dependence of the flip phenomenon on $(\minprim - \minsec)$ implies
that many systems will be affected by mass-loss as the primary becomes
a WD or a NS.  Indeed, many MS triple star
systems will have $\minprim \sim \minsec$, because the inner binaries of
solar-like triples typically have a flat mass distribution \citep{mazeh92,
raghavan10} and may even have a preference for ``twins'' \citep{pinsonneault06,
raghavan10}.  The similar masses of the inner binary suppresses the eccentric
Kozai mechanism since \epsoct{} is small.  

Furthermore, a significant
fraction of triple systems (those with modest mutual inclination) 
will have inner binaries that are wide enough to
avoid tidal contact and collisions on the MS, even if the system undergoes normal
Kozai-Lidov oscillations with $e_{\textrm{max}}$ given approximately by Equation
(\ref{eq:kozaiemax}).  However, once the primary of the inner binary evolves off
the MS and loses the majority of its mass, the system effectively
enters the test-particle approximation in many cases, particularly
for intermediate- and high-mass primaries where the change in mass
as the MS star becomes a compact object is large.  After mass loss,
\epsoct{} can increase dramatically, allowing the systems to exhibit the
eccentric Kozai mechanism, and driving the inner binary to extremely high
eccentricities and tidal contact or collision.  The likely result of the MIEK mechanism is
a close compact-object(WD/NS)-MS binary with separation $\sim \rsun$.  Such 
a configuration would produce a variety interesting astrophysical systems.

In this paper we demonstrate this "Mass-Loss Induced Eccentric Kozai" (MIEK) 
mechanism and provide a preliminary
exploration of its dependence on the initial eccentricities and mutual inclination of
the triple system, saving a detailed treatment for a future paper. In Section
\ref{sec:MassLoss}, we describe our method for integrating the orbits of triple
systems with mass-loss, and we give two example systems with different mass-loss
time-scales to illustrate the MIEK mechanism. By directly integrating triple
systems with an $N$-body code we circumvent the need to pick a limiting order when
expanding the three-body Hamiltonian and solving the secular dynamics.  In
Section \ref{sec:parameters}, we explore parameter space by investigating the
effects of eccentricity and inclination on the MIEK mechanism for an example
ternary.  We find that a fraction of widely separated triple systems
are brought to tidal contact for the first time only after the primary
evolves off the MS. In Section \ref{sec:discussion}, we discuss the 
possible outcomes and the implications of systems whose inner binary
comes to tidal contact or collides at each evolutionary phase.
In Section \ref{sec:conclusion},  we provide a brief conclusion.

\section{Simulating Mass-Loss in Triples}
\label{sec:MassLoss}

To simulate the dynamical effects of mass-loss on hierarchical triple systems,
we modified the $N$-body code FEWBODY\footnote{FEWBODY is now available at http://fewbody.sourceforge.net/.} \citep{fregeau04} to create triple systems
and evolve them. We create a triples with specified inner and outer semi-major axess (\ain and \aout), 
eccentricities (\ein and \eout), arguments of periastron (\gin and \gout), primary mass
in the central binary (\minprim), secondary mass in the central binary
(\minsec), tertiary mass (\mout), random phase angles, and mutual
inclination between the two orbits (\cosi). 

During integration of the triple, we check that the triple remains bound and stable.  If the
triple becomes unbound we continue integrating the system until a conservative minimum tidal
perturbation on the binary is met ($\delta < 10^{-7}$; see \citealp{fregeau04}), and we thus include any possible resonant interactions which
could result when the triple system becomes unstable\footnote{This allows our calculations to include systems affected by the ``triple evolution dynamical instability'' (TEDI, \citealp{perets12}).}.  Integration is also stopped
if the time steps become exceedingly small. This occurs for a handful of triples
where the separation between the inner binary approaches 0 because the stars are
on nearly radial orbits. However, in the current study these cases only
occur when the inner two stars are separated by $\ll \rsun$.

\begin{figure}
	\centerline{
		\includegraphics[width=9cm]{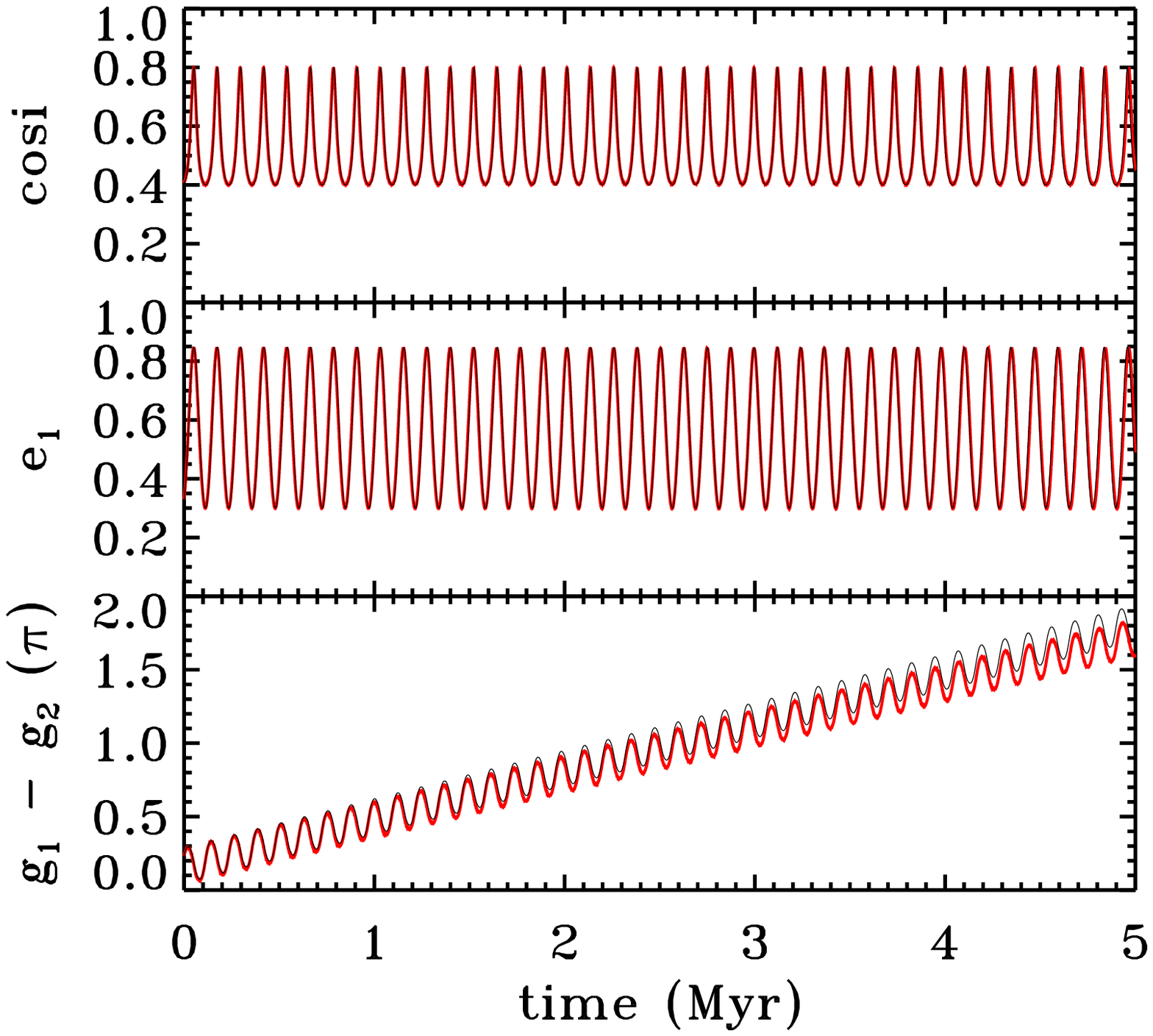}
	}
	\caption{Comparison between our modified FEWBODY code (black), and an octupole-order secular code (red, \citealp{blaes02}). The triple system has 
$\minprim = 7.0 \ \msun$, $\minsec = 6.5 \
\msun$, $\mout = 6 \ \msun$, $\ain = 10 \ \rm{AU}$, $\aout = 250 \ \rm{AU}$, $\ein =0.3$, $\eout = 0.1$, $\gin = 0\degree$, $\gout = 0\degree$, and $\cosi = 0.4$. There is a good agreement between the calculations but there is a slow drift in $\gin - \gout$.  Figure discussed in \S\ref{sec:MassLoss}.}
	\label{fig:Compare}
\end{figure}

To check our modified FEWBODY integrator we tested a sample of triple systems against the N-body
integrator MERCURY \citep{chambers97} and found good agreement.  We then tested our code against the octupole-order code of \citet{blaes02} with the general relativistic terms removed. An example triple system is shown in Figure \ref{fig:Compare}. As can be seen in the figure there is reasonable agreement with the secular octupole-order calculation.  There is, however, a slow drift in $\gin - \gout$ between the N-body codes and the secular code.  

\subsection{Evolutionary Phases}
\label{sec:Phases}

When we evolve a triple system, we break its evolutionary phases into four distinct stages: (1) primary on the MS, (2) primary evolved into a giant but prior 
to significant mass loss, (3) during a period of significant mass loss, and (4) after the 
end of mass loss when the primary has become a compact object. 
However, because of the prohibitive computational time required to integrate close and highly eccentric triples for the full
MS lifetime of the primary, we truncate our integrations after $10^{3}$ Kozai-Lidov times (\tk{}) defined by \citep{innanen97, holman97}:
\begin{eqnarray}
  \tk&=&\frac{4}{3}\left(\frac{\ain^3
  (\minprim + \minsec)}{G\mout^2}\right)^{1/2}\left(\frac{\bout}{\ain}\right)^3
 \nonumber \\
  &\simeq& 6.4 \times 10^{4} \textrm{yr} \left(\frac{\ain}{10
  \textrm{AU}}\right)^{3/2}  \left(\frac{\minprim + \minsec}{13.5
  \msun}\right)^{1/2}  \left(\frac{6\msun}{\mout}\right) 
  \left(\frac{1}{25}\frac{\bout}{\ain}\right)^{3} \nonumber \\
  \label{eq:tk}
\end{eqnarray}
where $\bout =\aout (1 - \eout)^{1/2}$. Thus, we implicitly assume that the triple systems have no interesting behavior after this time.  

We account for the MS lifetime of the primary by only integrating the triple systems for a time (\tms) defined by:
\begin{equation}
 \tms = \textrm{min}(\tmsprim, 10^{3} \tk) \times (0.95 + 0.1 \xi),
 \label{eq:tms}
\end{equation}
where $\xi$ is a uniformly distributed random number on the interval $[0,1)$,
\tmsprim{} is the MS lifetime of the primary.  
We determine the MS lifetimes
by interpolating the logarithmic lifetimes on a grid of stellar models run with
the Yale Rotating Evolution Code with input physics described in
\citet{vansaders11}.  We multiply by the second factor in Equation
(\ref{eq:tms}) in order to randomize exactly when the primary evolves off the main
sequence to avoid any correlations of the start of mass loss between different
triple systems in our sample. 

To simulate the giant phase of the primary, we continue to integrate the triple
for a time (\trg) defined by:
\begin{equation}
  \trg = \rm{min}(0.1 \tmsprim, 10^{3} \tk).
\end{equation}

Then, to simulate mass-loss, we linearly decrease \minprim{} over a mass-loss
time scale (\tml) to the WD mass as specified by the initial-final mass relation
of \citet{kalirai08} given by:
\begin{equation}
 m_{0,\textrm{f}} = (0.109 \pm 0.007) \, \minprim + (0.394 \pm 0.025) \, \msun,
 \label{eq:IFmass}
\end{equation}
where $m_{0,\textrm{f}}$ is final mass of the primary.  
To test our mass-loss prescription, we integrated binaries where one member underwent
mass loss and tested that the change in the semi-major axis and eccentricity of
the binary matched that for instantaneous mass loss and adiabatic mass loss
(i.e., Equations 4--6 in \citealp{baribault94})  when $\tml \ll P$ and when $\tml
\gg P$, respectively, where $P$ is the orbital period of the orbit.  We found
good agreement.

After mass-loss has
finished we integrate the evolved triple for a time (\tafter) given by:
\begin{equation}
 \tafter = \rm{min}(\tmssec - 1.1 \tmsprim - \tml, 10^{3} \tk),
\end{equation}
where \tmssec{} is the MS lifetime of the next most massive star in
the triple system and \tk{} is redefined by  Equation (\ref{eq:tk}) with the orbital
parameters of the triple right after the conclusion of mass-loss.

\subsection{Tidal Contact and/or Collisions}
\label{section:tidal_criterion}

A detailed treatment of stellar tides and tidal dissipation is beyond the scope of this paper.  However, to determine when tidal effects are likely to be important, we define an ad-hoc minimal separation between the inner two binary members, \rtide{}, below which strong tidal contact or a collision is assumed to have occurred.  That is, if the separation between the primary and secondary is $r < \rtide$, we assume either (1) that tidal dissipation will remove energy from the inner orbit decreasing \ain{} or (2) if the separation between the inner binary members changes dramatically in a single inner orbital period (\pin{}), a physical collision may occur before tides become important, as in the recent work of \citet{katz12}.  In either case, our simulations are not valid for $r < \rtide$.  

When choosing a tidal criterion we considered two tidal effects which will alter the evolution of the system.  The first is the nondissipative contribution that tidal bulges on the members of the inner binary have on apsidal motion \citep{sterne39}.  This additional apsidal motion may detune the Kozai mechanism (e.g. \citealp{wu03}; \citealp{fabrycky07}). We find that the timescale of this apsidal motion becomes comparable to \tk{} when $\rperi \sim 4 \Rinprim$, assuming the typical tidal Love number valid for $n=3$ polytropes, $k=0.028$, \citep{eggleton01} and the system parameters for the fiducial system presented in \S\ref{sec:MIEK}.  The second effect is eccentricity damping due to dissipative tides \citep{mazeh79, wu07}. For our fiducial system, we derive a critical inner seperation by equating the tidal and Kozai forcing terms in the equation for the secular evolution of \ein{}.  We find a critial inner separation of:
\begin{equation}
 r_{\textrm{crit}} \sim 3 \Rinprim \left( \frac{10^4}{Q} \ \frac{k}{0.028} \right)^{2/13},
 \label{eq:dissipative}
\end{equation}
where $Q$ is the tidal dissipation factor.  As can be seen in Equation (\ref{eq:dissipative}), $r_{\textrm{crit}}$ is very insensitive to the exact values chosen for $k$ and $Q$.  For simplicity, and to allow a broad, albeit incomplete, survey of parameter space, we take $\rtide = 3 \times \rm{max}(\Rinprim, \Rinsec)$ for the bulk of this study, but we investigate the effect of changing \rtide{} on our results in \S\ref{sec:rtide}. 

\subsection{General Relativistic Precession}
\label{sec:GRprecession}

General relativistic (GR) periastron precession can
``detune'' the normal Kozai
mechanism, and truncate the maximum eccentricity attainable for a
triple system \citep{blaes02, miller02, thompson11}. FEWBODY does not contain
post-Newtonian corrections, so we must check if it
is permissible to ignore GR precession.  The GR
precession timescale (e.g., Equation 23 of
\citealp{fabrycky07}) is given by:
\begin{eqnarray}
  \tgr&=&
  \frac{1}{3}\frac{\ain}{c}\left(\frac{\ain
  c^2}{G(\minprim + \minsec)}\right)^{3/2}\hspace*{-0.2cm}(1-\ein^2) \nonumber
\\
  &\simeq&3.4\times10^7{\rm \,\,yr}\,\,
  \left(\frac{13.5 \msun}{(\minprim +
  \minsec)}\right)^{3/2}\left(\frac{a_1}{10{\rm 
  AU}}\right)^{5/2}(1-e_1^2).\nonumber \\
  \label{eq:tgrp}
\end{eqnarray}
As discussed in \citet{blaes02}, the Kozai mechanism only operates if $\tk\lesssim\tgr$.  
We see that for the masses and semi-major axes of our fiducial case presented in \S\ref{sec:MIEK}, $\tk\lesssim\tgr$ for any choice of \eout{} until the inner eccentricity is increased such that $\rperi{} < \Rinsec$. Thus the normal Kozai mechanism is free to operate in our triple integrations.  

However, we also see in \S\ref{sec:MIEK} that
the eccentric Kozai timescale (\tek), the timescale for "flips" and extreme eccentricity
maxima, is much longer then \tk{}, $\tek \sim 10^6$
years. Furthermore, we see that $\tek > \tgr$ for values of \ein{} where $\rperi
> \rsun$, which would seem to imply that GR precession might detune the eccentric
Kozai mechanism before tidal effects become important.
This is not clear however, because the
majority of the GR
precession occurs when \ein{} is large, which is only a small fraction of
the eccentric Kozai cycle. To test if GR precession will suppress the
maximum \ein{} reached, we take the fiducial triple system after mass loss that exhibits the eccentric Kozai flip, and compare this to the equivalent
octupole-order calculation that includes GR precession using the code
of \citet{thompson11} \citep{blaes02}.  We found good
agreement between the two calculations, implying that GR precession
does not shut off the eccentric Kozai mechanism, and is thus not important for the 
current study.  However, these effects should be more carefully considered in
future studies where \ain{} is much smaller, and when the periastron distance 
can be significantly smaller than $R_\odot$ if both the primary and secondary are compact objects.

\section{MIEK: \\The Mass-Loss Induced Eccentric Kozai Mechanism}
\label{sec:MIEK}

\begin{figure*}[htp]
	\centerline{
		\includegraphics[height=8.5cm, angle=90]{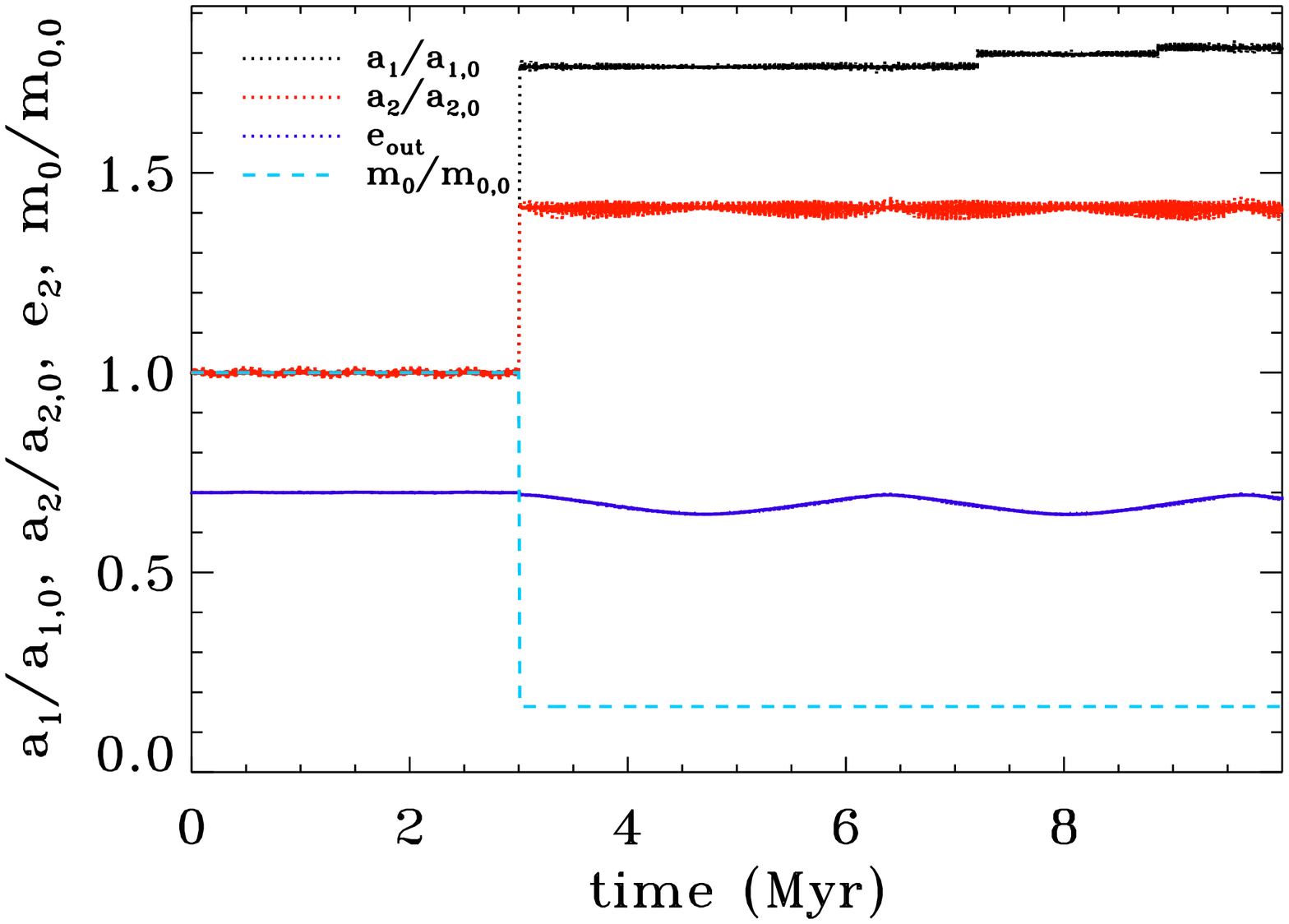}
		\includegraphics[height=8.5cm, angle=90]{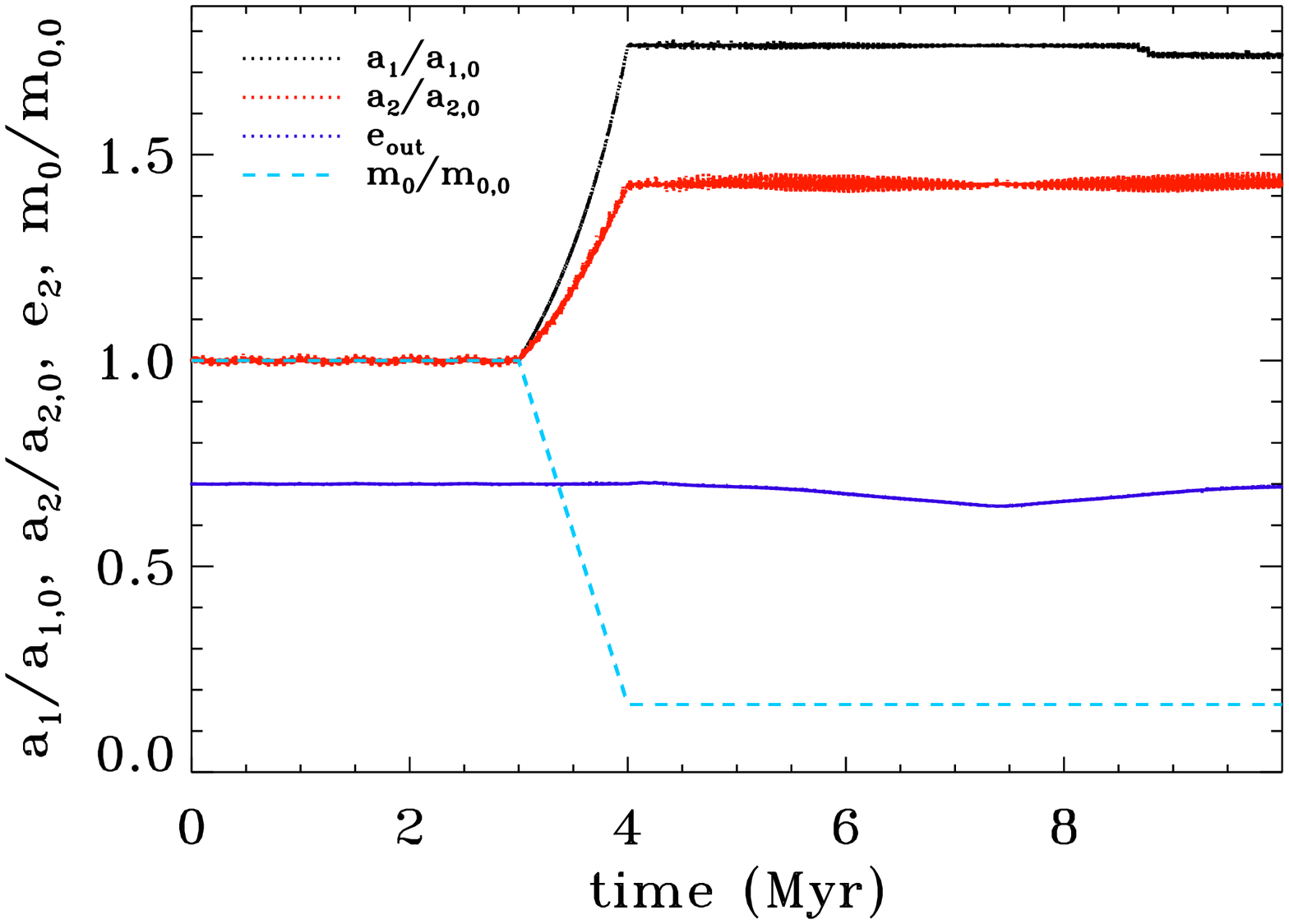}
	}
	\centerline{	
		\includegraphics[height=8.5cm, angle=90]{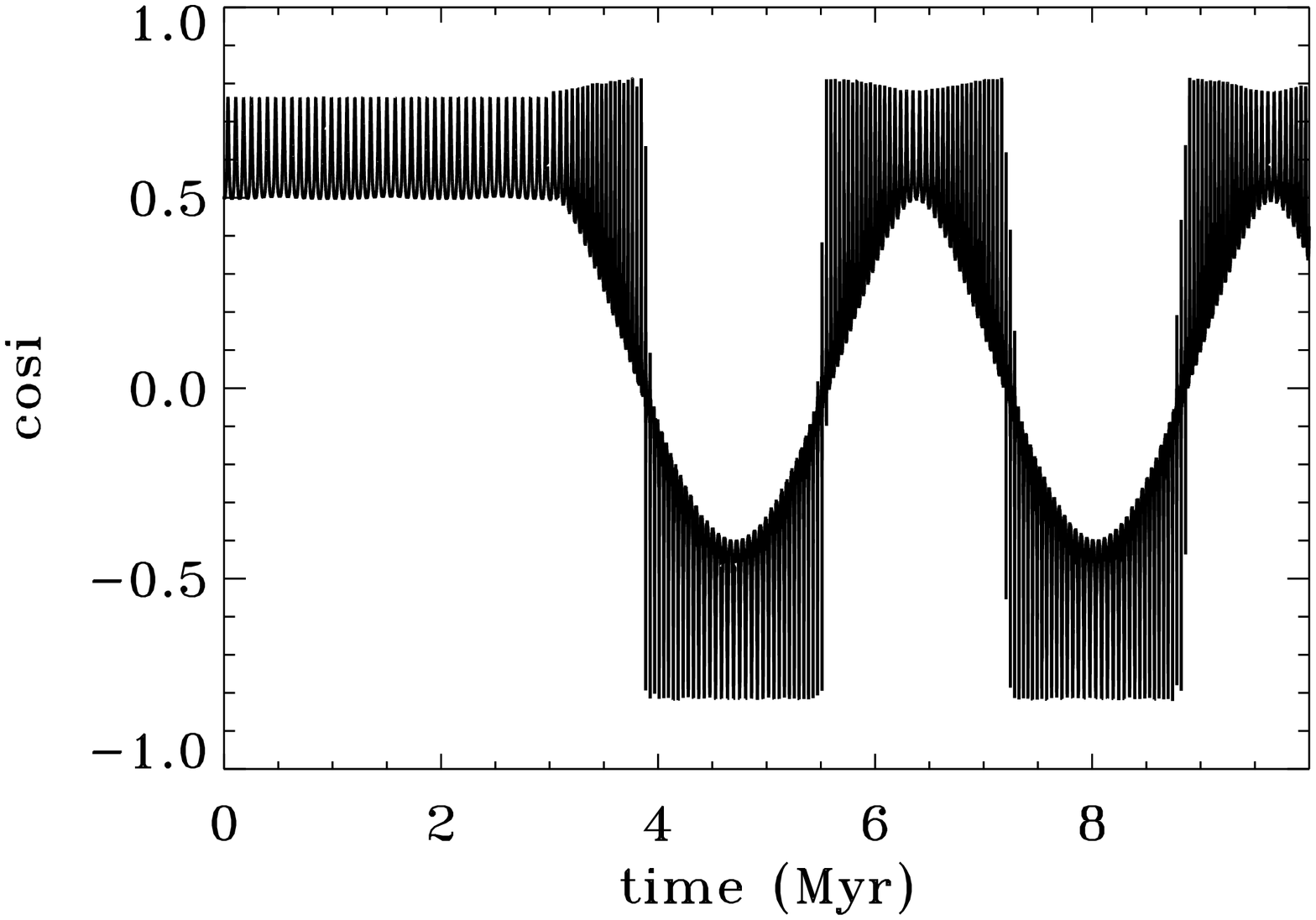}
		\includegraphics[height=8.5cm, angle=90]{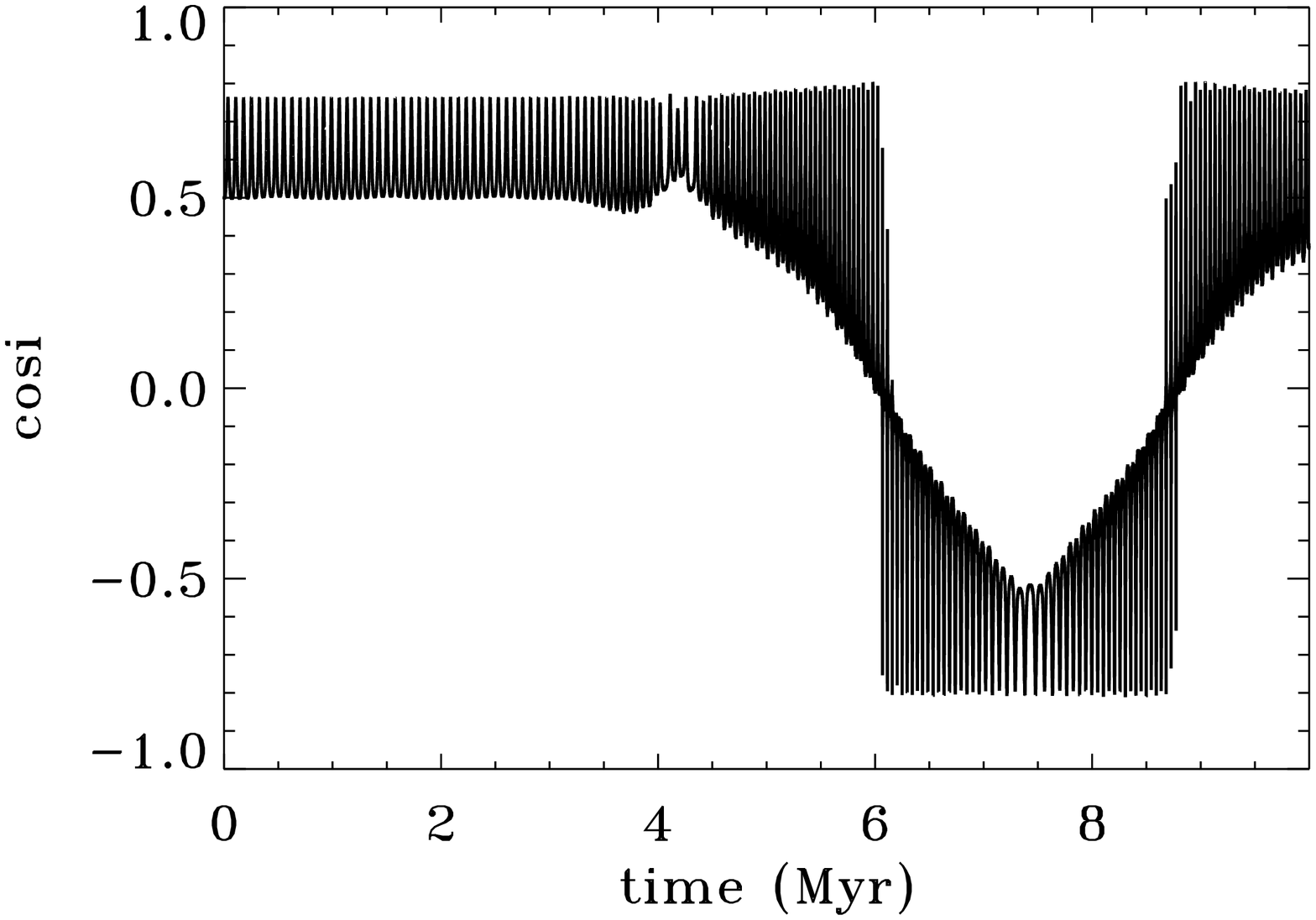}
	}
	\centerline{
		\includegraphics[height=8.5cm, angle=90]{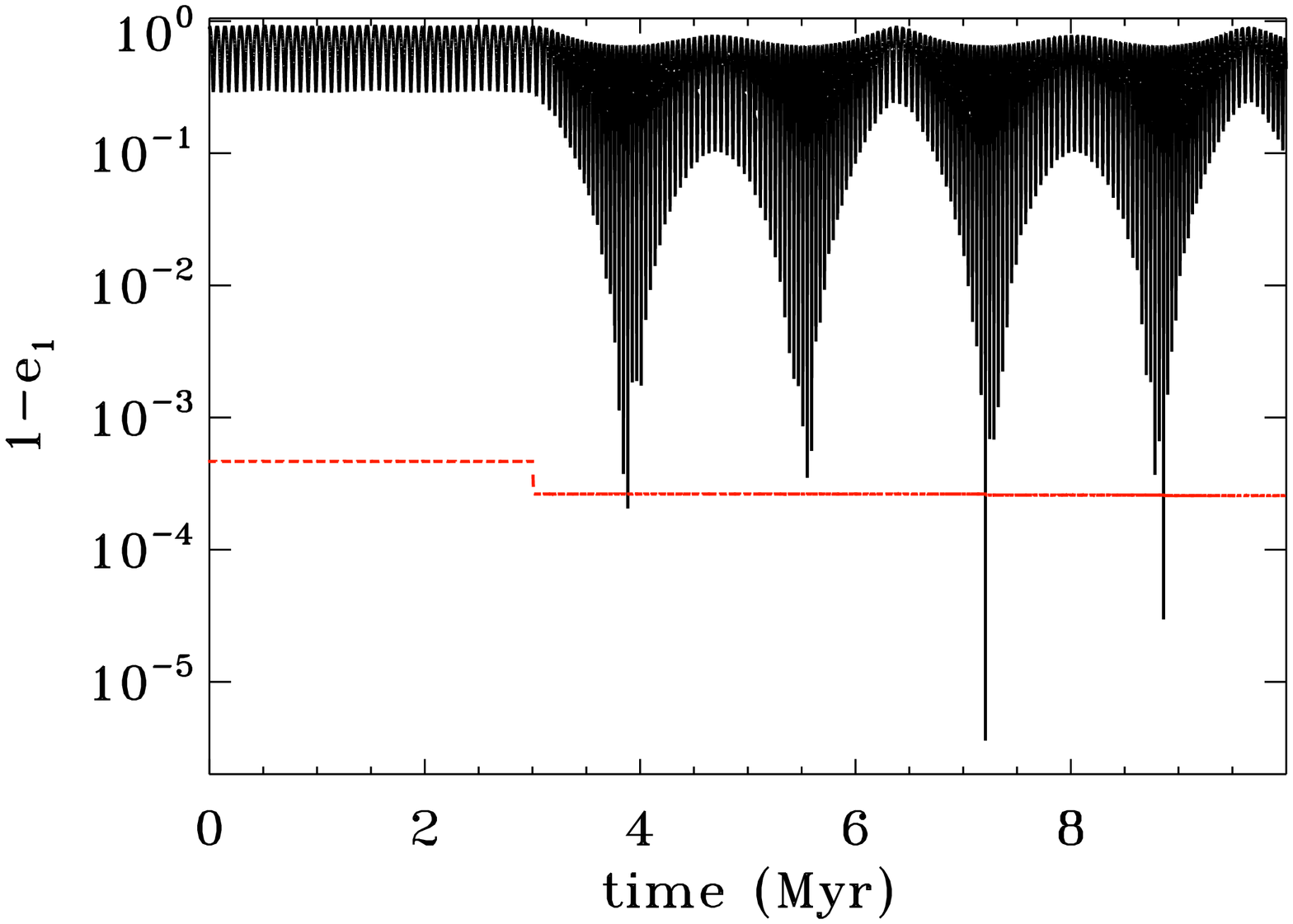}
		\includegraphics[height=8.5cm, angle=90]{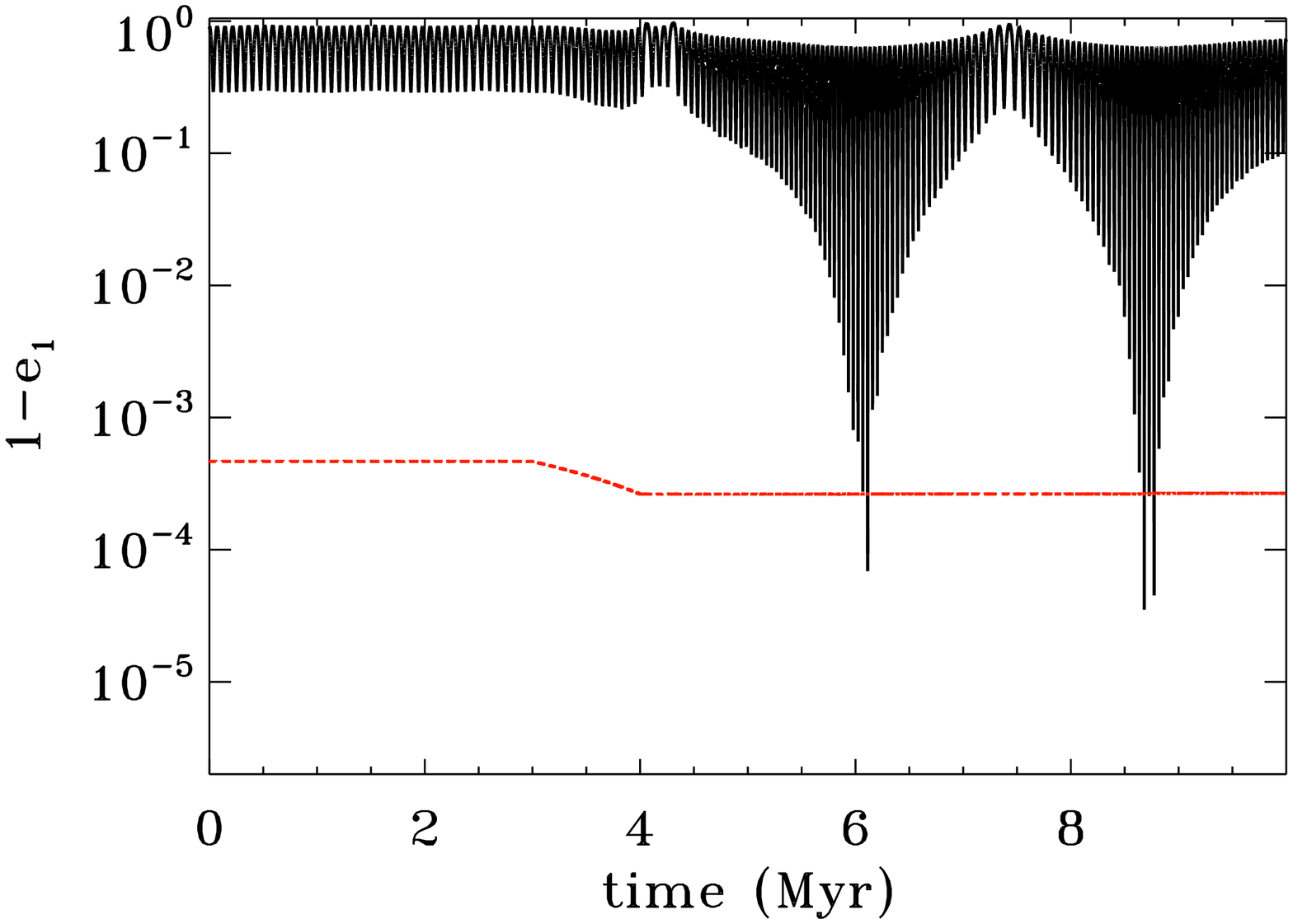}
	}
	\caption{The ``Mass-loss Induced Eccentric Kozai'' (MIEK) Mechanism. Two example systems with the left and right panels having
mass-loss timescales of $\tml = 10^{4} \rm{yr}$ and $\tml = 10^{6}
\rm{yr}$, respectively.  Triples have $\minprim = 7.0 \ \msun$, $\minsec = 6.5 \
\msun$, $\mout = 6 \ \msun$, $\ain = 10 \ \rm{AU}$, $\aout = 250 \ \rm{AU}$, $\ein =
0.1$, $\eout = 0.7$, $\gin = 0\degree$, $\gout = 180\degree$, and $\cosi = 0.5$.
{\it Top Panels:} Evolution of \ain{} (dotted black), \aout{} (red dotted),
\eout{} (blue dotted), and \minprim{} (light blue dashed).  {\it Middle Panels:}
Time evolution of \cosi. {\it Bottom Panels:} Time evolution of \ein. Note the qualitative difference between the normal
Kozai-Lidov mechanism ($\ein \rightarrow 0.7$) before the onset of mass loss at 3 Myr and the eccentric Kozai
mechanism ($\ein \rightarrow 1$) after mass loss when $\rperi < \rsun$ (denoted by the red dashed line) and flips occur quasi-periodically.  We caution that our integrations after the first
eccentricity maximum are for purposes of illustration only since we do not include
the effects of tides.}
	\label{fig:FigExampleSystems}
\end{figure*}

Figure \ref{fig:FigExampleSystems} presents two example systems with the left
and right panels having mass-loss time scales of $\tml = 10^{4}$ yr and $\tml
= 10^{6}$ yr, respectively.  Both have $\minprim = 7.0 \ \msun$, $\minsec = 6.5
\ \msun$, $\mout = 6 \ \msun$, $\ain = 10 \ \rm{AU}$, $\aout = 250 \ \rm{AU}$, $\ein =
0.1$, $\eout = 0.7$, $\gin = 0\degree$, $\gout = 180\degree$, and $\cosi = 0.5$
($i=60 \degree$).  The top panels present the evolution of \ain, \aout, \eout, and
\minprim.  The middle and bottom panels show the evolution of \cosi{} and $1 -
\ein$, respectively.  In the bottom panel, the red line shows where \ein{} is
high enough such that the radius of periastron of the inner binary (\rperi) is
equal to \rsun.  The example systems go through normal Kozai-Lidov cycles before
mass-loss begins, reaching a maximum $\ein \sim 0.75$ as predicted by Equation
(\ref{eq:kozaiemax}), but \rperi{} is still more than two orders of magnitude
too large for tidal interactions to be significant. 

After adiabatic mass-loss $\minprim
\rightarrow 1.15 \msun$, and \epsoct{} increases (Equation \ref{eq:epsoct}).  To calculate this
increase analytically, we first note that the increase in the semi-major axis
and change in eccentricity of a binary that underwent adiabatic mass-loss is
given by:
\begin{equation}
 \frac{a_{\textrm{f}}}{a_{0}} = \frac{M_{0}}{M_{\textrm{f}}}, \; \;  \textrm{and} \; \;  e_{\textrm{f}} = e_{0},
 \label{eq:adiabatic}
\end{equation}
where $M_{0}$ and $M_{\textrm{f}}$ are the initial total mass and final total mass of
the binary, respectively.  From Equation (\ref{eq:adiabatic}) we see that the
change in the ratio of the semi-axes of a triple system that underwent adiabatic
mass-loss is
given by:
\begin{equation}
 \frac{\aoutf}{\ainf} =
\frac{\aouti}{\aini} \left( \frac{\minprim +
\minsec + \mout}{\minprimf + \minsec + \mout} \right) 
\left( \frac{\minprimf + \minsec}{\minprim + \minsec} \right),
 \label{eq:adiabaticT}
\end{equation}
where \aouti{} and \aoutf{} are the initial and final outer semi-major axes,
respectively, and \aouti{} and \aoutf{} are the initial and final inner
semi-major axes, respectively. From Equation (\ref{eq:adiabaticT}) one sees that the outer to inner semi-major axis ratio decreases for our fiducial system during adiabatic mass loss by $\aoutf / \ainf \simeq 0.8 \ \aouti / \aini$.   Then with Equation (\ref{eq:epsoct}) we see
that the fractional increase\footnote{While
\epsocti{} and \epsoctf{} will vary between systems in
\S\ref{sec:parameters} depending on \eout, the fractional increase of \epsoct{}
will, however, be the same because the mass loss is adiabatic so \eout{} drops out.} in
\epsoct{} due to adiabatic mass loss of the primary is given
by:
\begin{equation}
 \frac{\epsoctf}{\epsocti} =
\left|
\left( \frac{\minprimf - \minsec}{\minprim - \minsec} \right)
\left( \frac{\minprimf + \minsec + \mout}{\minprim + \minsec + \mout} \right)
\left( \frac{\minprim + \minsec}{\minprimf + \minsec} \right)^{2}
\right|,
 \label{eq:adiabaticEpsOct}
\end{equation}
where  \epsocti{} and \epsoctf{} are the initial and final values of \epsoct, respectively. The absolute value is due to the ambiguity as to which of the
binary stars
will be the most massive in the final system.  After mass loss our
fiducial
triple system is in the near-test-particle limit and \epsoct{} increases by a
factor of \EpsOctIncrease, from \EpsOctOrig{} to \EpsOctFinal.  These
factors enable
the eccentric Kozai mechanism, the system then flips, and as \cosi{} passes through
$0$, $(1 - \ein)$ approaches $10^{-4}$. Both systems would be affected by tidal
interactions during their first flip as $\rperi < \rsun$.  This example illustrates the essence of the MIEK mechanism.

\section{Inclination and Eccentricity Dependence}
\label{sec:parameters}

The importance of the MIEK mechanism for a full population of triple systems
will depend on the adopted mass function of the stars, the
eccentricity and semi-major axis distributions of the inner binary and outer
tertiary, as well as the mutual inclination distribution.  We save a full
exploration of parameter space for a future paper, but we explore the effects of
inclination and eccentricity on the dynamics of the example system from
\S\ref{sec:MIEK}. This system is an interesting test case because all the
masses are approximately equal, so the octupole-order terms mostly become important
during and after mass loss.  

\subsection{Parameters Space Exploration}
\label{sec:grid}

To explore the effects of inclination and eccentricity, we integrate \NumTriples{} triple systems over an equally spaced grid
with \Numcosi, \Numein, \Numeout, and  \Numgin{} values of \cosi, \ein, \eout,
and \gin, respectively, while keeping the initial masses and initial semi-major axes from our example
system.  Because the outcomes of triple systems effected by MIEK may differ depending 
on the initial phase angles, we integrate three triple systems at each grid point in 
\cosi, \ein, \eout, and \gin with randomized initial phase angles.

In order to determine if the inner binary of each system becomes tidally affected or collides by our ad-hoc tidal criterion (Section \ref{section:tidal_criterion}),
 we record the minimum inner separation (\rmin) in each of the four evolutionary phases described in \S \ref{sec:MassLoss}.
We integrate these triple systems assuming $\tml = 10^{4}$ years and posit that
tides will affect the inner binary if \rmin{} is smaller than some ad-hoc tidal radius
(\rtide).  We define this ad-hoc tidal radius differently for the different evolutionary phases of the primary, in an attempt to capture the different strength of tidal forces due to the varying physical size and mass of the primary.  Before the primary evolves off the MS we take $\rtide = \TooCloseMS$, where \Rinprim{} in the MS radii of the primary.  When the primary is a giant and during the mass-loss phase, we take $\rtide = \TooCloseRG$. Finally, when the primary becomes a WD we take $\rtide = \TooCloseWD$, where \Rinsec{} in the MS radii of the secondary in the inner binary.  We determine \Rinprim{} and \Rinsec{} from the \citet{torres10} sample by taking the median of the observed radii for stars with similar masses ($\left| \log{M} - \log{M'} \right| < 0.1$).

Additionally, we would like to determine when the eccentric Kozai mechanism is operating. Because orbital flips (i.e., a change of sign in \cosi) only occur during the eccentric Kozai mechanism, we monitor each triple systems for flips during each of the aforementioned evolutionary phases. We then use the two above criteria to loosely define systems that are strongly affected by the MIEK mechanism, as systems that come into tidal contact and exhibit a flip after the onset of mass loss.

\subsection{Results}
\label{sec:Results}

\begin{figure*}[]

	\centerline{
		\includegraphics[width=8.5cm, angle=90]{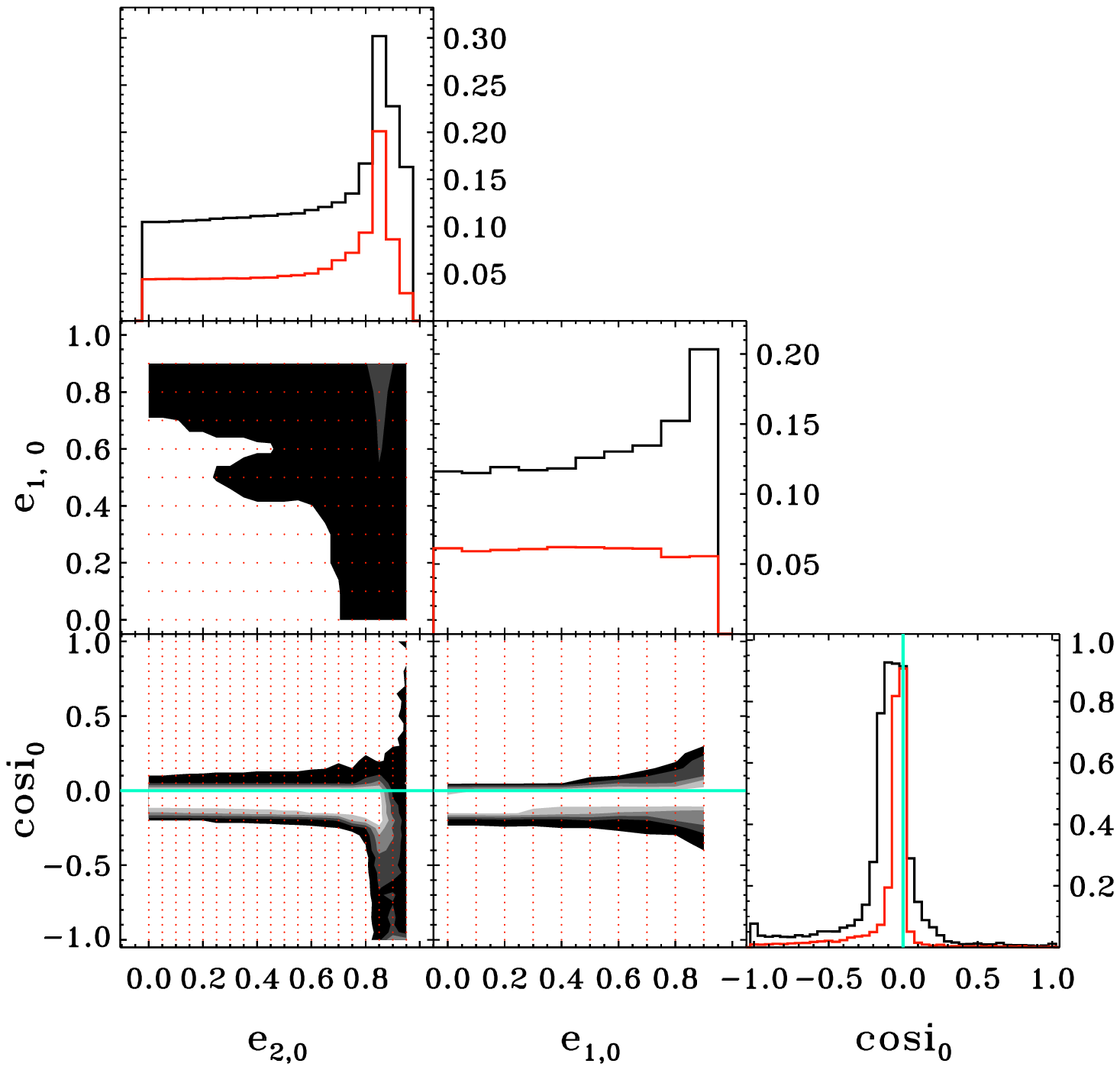}
		\includegraphics[width=8.5cm, angle=90]{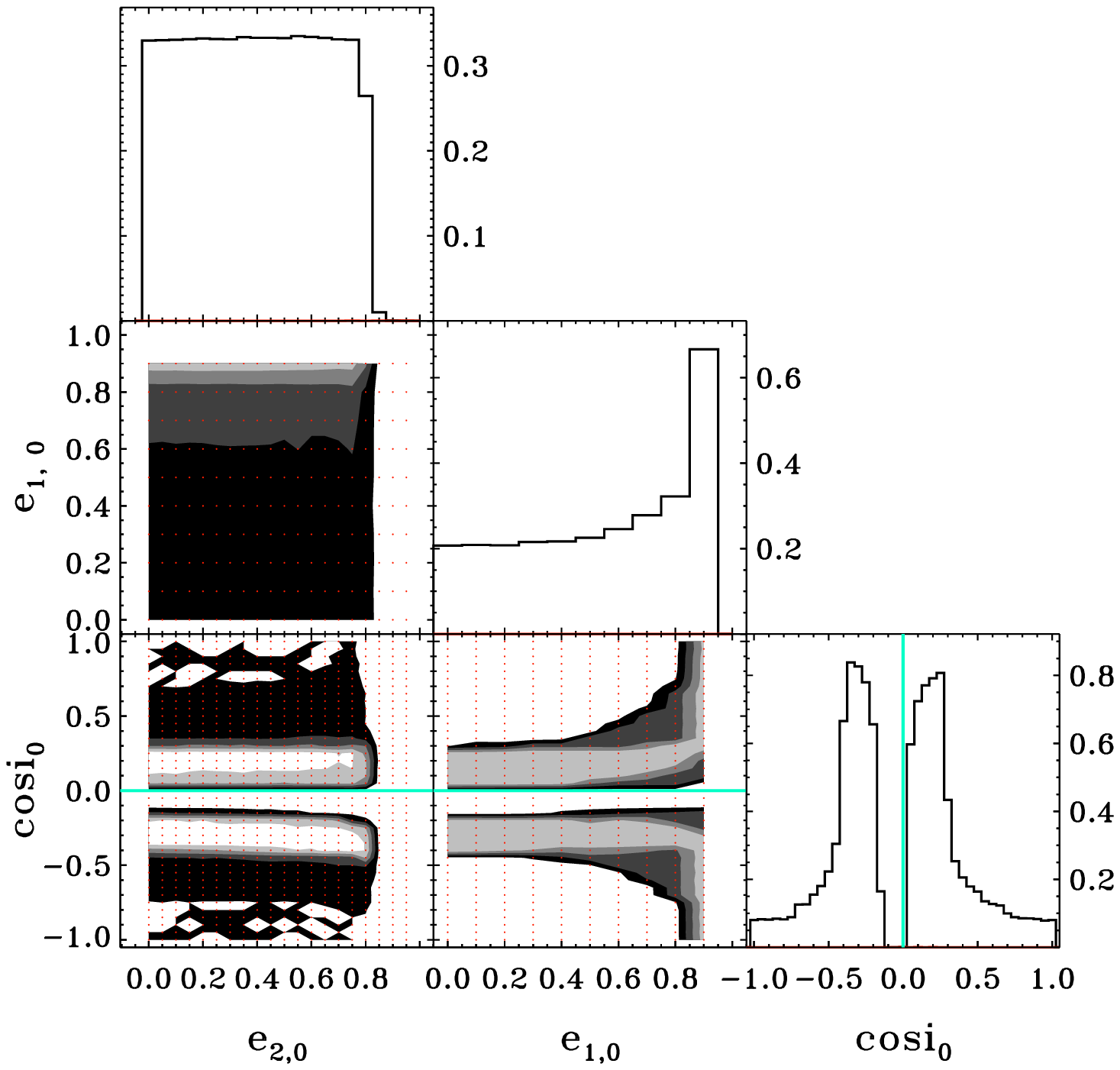}
	}
	\caption{Initial parameters for triple systems that were tidally
affected or collided on the MS, $\rperi < 3 \Rinprim$ (left), and triple
systems that were tidally affected or collided after the primary evolved off the main
sequence but before the onset of significant mass-loss, $\rperi < 1 \textrm{AU}$ (right). 
In the histograms, black shows the fraction of systems in each bin that are
tidally affected or collide and red shows the fraction of systems that also exhibited an
orbital flip.  Contours present the fraction of systems at each grid point that
would be tidally affected by our criteria.  The contours are in steps of 20\%
starting at 10 \% (black) Red dots show the grid points where systems
where integrated. The teal line on the contours show where $\cosi = 0$.  Note
that there is a pronounced asymmetry between prograde and retrograde triples. Figure discussed in
\S\ref{sec:parameters}.}
	\label{fig:FigPar}
\end{figure*}

\begin{figure*}[htp]

	\centerline{
		\includegraphics[width=15cm, angle=90]{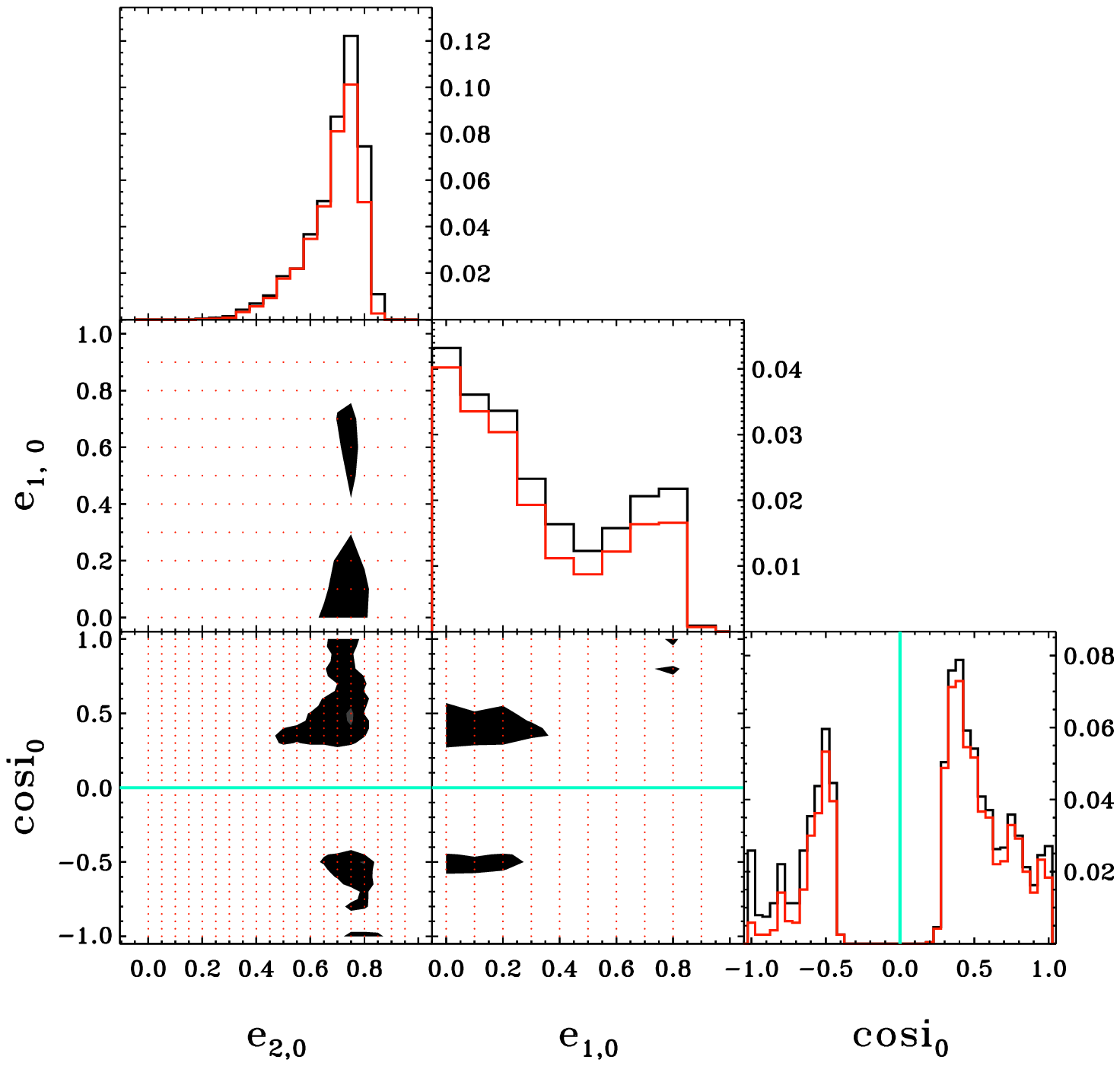}
	}
	\caption{Initial parameters for triple systems that were tidally
affected or collide after the end of mass loss, $\rperi < 3 \Rinsec$.  In the histograms, black shows the
fraction of systems in each bin that are tidally affected or collide and red shows the
fraction of systems that also exhibited an orbital flip.  Contours present the
fraction of systems at each grid point that would be tidally affected by our
criteria.  The contours are in steps of 20\% starting at 10\% (black). Red dots
show the grid points where systems where integrated. The teal line on the
contours show where $\cosi = 0$.  Note there is a pronounced asymmetry between prograde and
retrograde triples. Figure discussed in \S\ref{sec:parameters}.}
	\label{fig:FigParAfter}
\end{figure*}

The results of the parameter space search in inclination and eccentricity are presented in Table \ref{tab:ParmsOutcomes}, Figure \ref{fig:FigPar}, and Figure \ref{fig:FigParAfter}.  In Table \ref{tab:ParmsOutcomes} we break down the fraction
of the triple systems that flip and become tidally affected or collide during the various evolutionary stages presented in \S \ref{sec:MassLoss}.  We highlight any asymmetries about $\cosii = 0$ by presenting the fractions for prograde and retrograde systems separately. We now describe each of the evolutionary phases of the primary in turn.  

\input{tab1.tex}

When the primary is on the MS (``Primary MS" in Table \ref{tab:ParmsOutcomes}), \TidalBeforePer{} of triple systems become tidally affected or collide ($\rmin \leq \rtide$).  We show the initial inclinations and eccentricities of these triple systems in the left
panel of Figure \ref{fig:FigPar}.  Each contour plot displays the parameter space projected onto a two parameter plane. Each histogram marginalizes over all parameters except one, and then displays the fraction of systems in each bin that become tidally affected or collide. From the left panel of Figure \ref{fig:FigPar}, we see that when the primary is on the MS
the majority of systems which become tidally affected or collide are initially highly
inclined.  These systems undergo normal Kozai-Lidov oscillations, which
bring the inner binary to tidal contact with a maximum pericenter distance given by: 
\begin{equation}
 \rperi = \ain(1-e_{\rm{1, max}})=\ain\left[1- \left( 1 - \frac{5}{3} \cos^{2}i_0 \right)^{1/2} \right]
 \label{eq:kozairperi}
\end{equation}
with $e_{\textrm{1, max}}$ given by Equation (\ref{eq:kozaiemax}) and $i_0$ the initial eccentricity.  
Then, setting the periastron distance equal to $r_{\rm tide}$,
\begin{equation}
 \rperi = \rtide = 3 \Rinprim \sim 12 \rsun \sim 0.05 \textrm{AU} \sim 0.005 \ \ain,
 \label{eq:kozairperi2}
\end{equation}
we see that the required critical initial inclination to interact tidally when the primary is on the MS is
\begin{equation}
 \cos i_{0,\rm crit}\sim 0.08,
 \label{eq:kozairperi3}
\end{equation}
which is in good agreement with the prograde orbits in the left panel of Figure \ref{fig:FigPar}. Note, however, that the distribution is asymmetric about
$\cosii = 0$, with \RetTidalBeforePer{} of the retrograde systems becoming tidally affected or colliding while only \ProTidalBeforePer{} of the prograde systems do. This asymmetry is not captured by Equation (\ref{eq:kozaiemax}) and Equation (\ref{eq:kozairperi}), because of the approximations under which Equation (\ref{eq:kozaiemax})  is derived.  A more complete treatment of the quadrupole-order Hamiltonian shows that $e_{1,\rm max}$ does not occur at $\cos i_0=0$, but instead at retrograde mutual inclination (see Fig.~1 of \citealp{miller02}). 

Importantly, we find that \TidalFlipBeforePer{} of the triple systems, with $\cosii{} \simeq 0$, become tidally affected and exhibit an orbital flip, which is characteristic of the eccentric Kozai mechanism.  These systems comprise almost half of the total number of systems that become tidally affected or collide when the primary is on the MS, showing that the eccentric Kozai mechanism is also important. The implications that the eccentric Kozai has for previous studies involving the Kozai-Lidov mechanism and tidal friction is discussed 
further in \S \ref{sec:discussion}, but here we note that the systems that tidally interact or collide on the MS are analogous to the systems 
investigated by \citet{fabrycky07}, except that these authors only included quadrupole-order terms in the 
secular Hamiltonian, and thus may have missed the systems that undergo the eccentric Kozai
mechanism and execute a "flip" with $e_1\rightarrow1$ \citep{naoz11a, naoz11b}.

Additionally, \DistruptedBeforePer{} of triple systems become unbound when the primary is on the MS.  The systems that are unbound have large \eout{} and do not satisfy the standard, empirically derived, stability criterion for triple systems given by \citep{mardling01}: 
\begin{equation}
  \frac{\aout}{\ain} \geq  C \: f \left[ \left( 1 + \frac{\mout}{\minprim + \minsec} \right) \frac{1 + \eout}{(1 - \eout)^{3}} \right] ^{0.4},
 \label{eq:mardling}
\end{equation}
where $C=2.8$ and $f=1-\frac{0.3}{\pi}i$ is an ad-hoc inclination term.  
These systems are present in our study because we uniformly sample the eccentricities 
without regard for the initial dynamical stability of the system.
We see from Equation (\ref{eq:mardling}) and our initial assumed semi-major axes, that if $\cosi = 0$ then $\eout \leq 0.83$ must hold for the system to satisfy the stability criterion. It is interesting to note that in the \cosii{} vs. \eouti{} contour plot in the left panel of Figure \ref{fig:FigPar}, many triple systems which are unstable given the stability criteria become tidally affected or collide before they are unbound.  It is possible that tidal friction will quickly shrink \ain{} of these systems, causing them to become stable as the ratio $\aout/\ain$ increases.

After the primary has evolved off the MS but before significant mass-loss has begun
(``Primary Giant" in Table \ref{tab:ParmsOutcomes}), \TidalEvolvePer{} of triple systems become tidally affected or collide ($\rmin \leq 1 \rm{AU}$) for the first time.   We show the initial inclinations and eccentricities of these triple systems in the right panel of Figure \ref{fig:FigPar}.  Such a large fraction of the triple systems become tidally affected or collide during this stage of evolution because the physical size of the primary dramatically increases, so we impose an ad-hoc increase to  $\rtide = 1 \rm{AU}$.  Then from Equation (\ref{eq:kozairperi}) we find that 
\begin{equation}
\cos i_{0,\,\rm crit} \simeq 0.34,
 \label{eq:kozairperi3Giant}
\end{equation}
encompassing about a third of our parameter space.  Additionally, we see that only a small fraction of the systems exhibit the orbital flip characteristic of the eccentric Kozai mechanism when the primary is a giant, showing that the eccentric Kozai mechanism is not important during this phase of the primary's evolution.  Lastly, only \DistruptedEvolvePer{} of the systems are unbound during this evolutionary phase, emphasizing that initially dynamically unstable systems are either quickly unbound, when the primary is still on the MS, collide, or become tidally affected.

During the $\tml=10^{4}$ year mass-loss phase (``During Mass-Loss" in Table \ref{tab:ParmsOutcomes}) only a small fraction,
\TidalDuringPer{}, of the triple systems become tidally affected or collide for the first time.
This is because previous stages of evolution have removed the majority of triple systems that would tidally interact or collide with $\rtide = 1 \rm{AU}$, and because $\tml < t_{\rm{K}}$, so the systems do not have time to complete a Kozai cycle during the primary's mass loss episode.  We find that the fraction of systems which are unbound increases, to \DistruptedDuringPer{}, from the previous phase of evolution.  More systems become unstable because, as shown by Equation (\ref{eq:adiabaticT}), the outer to inner semi-major axis ratio decreases during mass loss.  Then, as a result of Equation (\ref{eq:mardling}), this decrease in the 
semi-major axis ratio causes more triple systems to be unstable.

Finally, after the end of mass-loss (``After Mass-Loss" in Table \ref{tab:ParmsOutcomes}) we see that \TidalAfterPer{} of triple
systems spanning a broad range in orbital parameters become tidally affected or collide for the first time.  We
note that the lion's share, \TidalFlipAfterPer{},  of these systems exhibit an orbital flip characteristic
of the eccentric Kozai mechanism and thus the MIEK mechanism because it is induced by mass loss of the primary. We now discuss the distribution of tidally affected or colliding systems after mass loss for the inclination and eccentricities individually.

The distribution of these tidally affected or colliding systems is asymmetric about $\cosii =0$, with \ProTidalAfterPer{} of prograde systems and only \RetTidalAfterPer{} of retrograde systems becoming tidally affect of colliding.  At least some of this asymmetry arises because, like previous stages of evolution of the primary, systems with $\cosii$ near $0$ are strongly preferred. However, earlier stages of evolution have removed a disproportionate fraction of the prograde systems, \ProTidalAfterPriorPer{}, compared to the fraction of retrograde systems, \RetTidalAfterPriorPer{}.  This causes the reverse asymmetry, more prograde than retrograde systems, in the distribution of systems which become tidally affected or collide after the end of mass loss.

The distribution in \eouti{} for the systems that become tidally affected or collide after the end of mass loss shows a strong preference for larger values, peaking at $\eouti{} = 0.75$.  This preference is easily understood when considering Equation (\ref{eq:epsoct}), which also shows a strong dependence on \eouti{}. Thus, the octupole order terms and consequently the eccentric Kozai mechanism are expected to become more important with larger \eouti{}.  Even though there is a strong preference for large \eouti{}, the MIEK mechanism continues to operate for systems with modest $\eouti \sim 0.4$. We also note that the \eouti{} distribution is truncated at very large values, which arises because most of these systems became tidally affected, collided, or were unbound when the primary was on the MS. 

Lastly, the distribution of \eini{} for the systems that become tidally affected or collide shows a preference for both large and small values.  The minimum in the distribution occurs at $\eini \sim 0.5$, with $\sim3$ times fewer systems becoming tidally affected or colliding then at $\eini = 0$. The preference for large values of \eini{} is easily understood because these systems do not need to increase their inner eccentricity as much to collide or come into tidal contact.  The preference for smaller values of \eini{}, however, is not easily understood.  We also note that the majority of systems with $\eini = 0.9$ become tidally affected or collide in earlier stages of the primary's evolution, so these systems are absent in the current \eini{} distribution.

\subsection{Sensitivity to \rtide{}}
\label{sec:rtide}

We now briefly evaluate the sensitivity of the results presented in \S \ref{sec:Results} on the ad-hoc tidal criterion chosen (Section \ref{section:tidal_criterion}).  

First, we evaluate the dependence of the fraction of systems that become tidally affected or collide on our choice of tidal criterion. We both decreased and increased the tidal criterion from $ \rtide = 3 \ R_{*}$ to $ \rtide = 2.5 \ R_{*}$, $ \rtide = 3.5 \ R_{*}$, $ \rtide = 5 \ R_{*}$, and $ \rtide = 10.0 \ R_{*}$.  $R_{*} = \Rinprim{}$ when the primary is on the MS and $R_{*} = \Rinsec{}$ when the primary is a WD. The fraction of systems that become tidally affected or collide after the end of mass-loss with these new tidal criteria are \TidalAfterRtideless{}, \TidalAfterRtidemore{}, \TidalAfterRtidefive{}, and \TidalAfterRtideten{}, respectively. When these fractions are compared to that presented in Table \ref{tab:ParmsOutcomes}, we see that the fraction of systems affected by the MIEK mechanism is relatively insensitive to the exact choice of tidal criterion.  This insensitivity results because an increase in the tidal radius on the MS affects systems that will already be affected by the large tidal radius when the primary 
becomes a Giant, and does not affect regions of parameter space shown in Figure \ref{fig:FigParAfter} where 
the MIEK mechanism operates.  In fact, increasing the tidal radius marginally increases the fraction of systems affected by MIEK. 

Additionally, we emphasize that by eliminating all systems that collide or were tidally affected
at a prior stage of evolution, \TidalFlipAfterPer{} is a conservative estimate of
the fraction of triple systems that will undergo the MIEK mechanism.  This is because many of the systems we remove from the sample as the evolutionary stages progress may still be viable candidates for the MIEK mechanism. One possibility is that tides on the giant primary only detune the normal Kozai-Lidov mechanism, and do not allow \ein{} to increase so dramatically.  Then, our removal of these systems from the sample would unrealistically decrease the number of systems that experience MIEK.  These systems would survive in our sample to the stage after mass loss is complete.  Another possibility is that the combination of the Kozai-Lidov mechanism and tides on a giant will work to shrink \ain{} to a few AU.  These systems will then have $\aout/\ain \sim 100$, but they can still be affected by the eccentric Kozai mechanism after mass loss, which will shrink this semi-major axis ratio according to Equation (\ref{eq:adiabaticT}).  

\begin{figure}[t]
	\centerline{
		\includegraphics[width=8cm, angle=90]{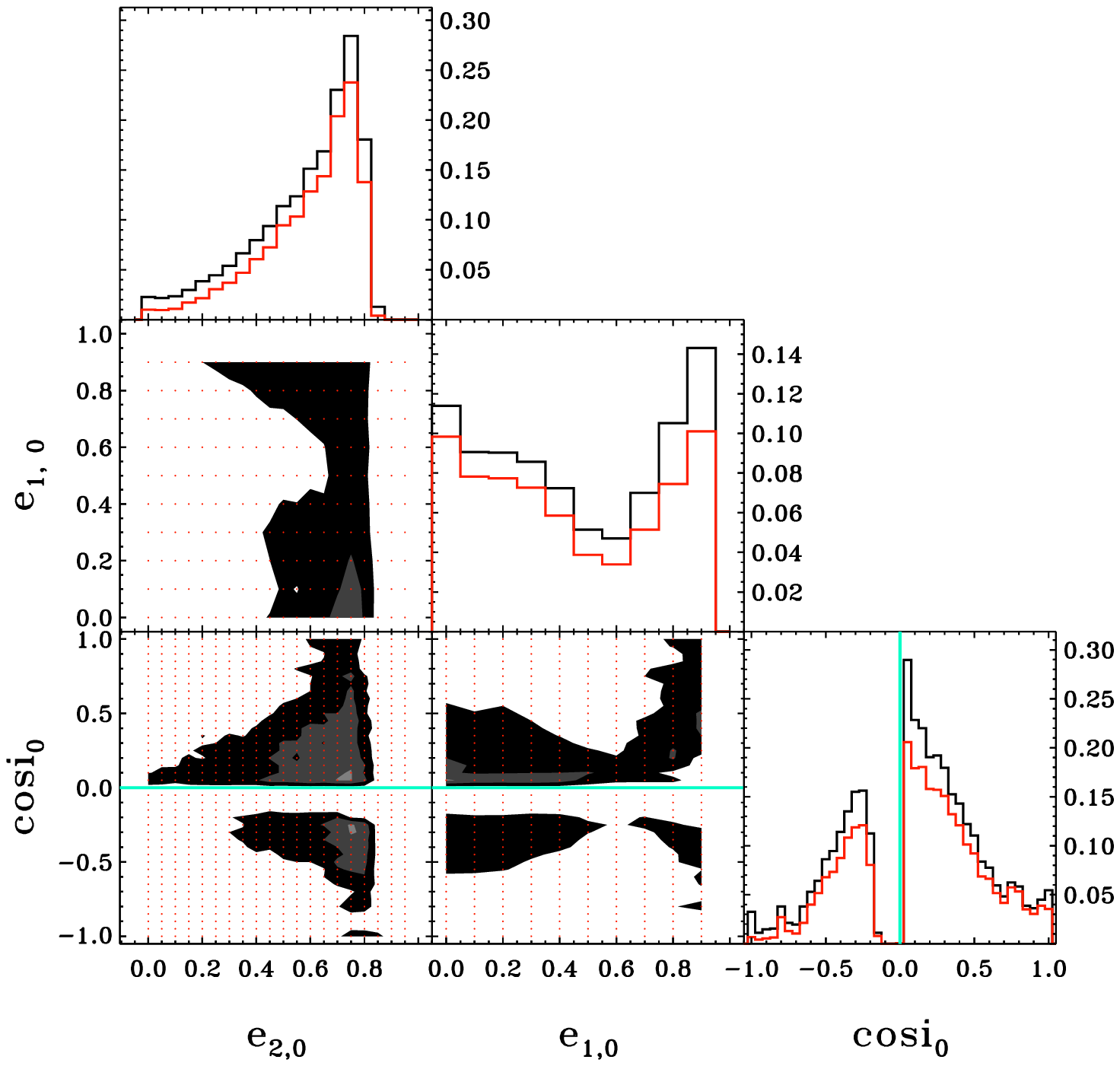}
	}
	\caption{Same as Figure \ref{fig:FigParAfter}, except assuming the
primary's tidal radius remains unchanged as it evolves until it becomes a
WD. This is taken to be an
upper bound on the fraction of systems that become tidally affected or collide after the
end of mass loss.  Figure discussed in \S\ref{sec:parameters}.}
	\label{fig:FigParAfternoRRG}
\end{figure}

Evaluating these effects would involve a realistic calculation of the 3-body dynamics with tides since
$\cos i$ will evolve secularly during the tidal interaction \citep{fabrycky07}.  We
save this for a future work.  Here, 
to roughly quantify the number of systems that could potentially be affected by the MIEK mechanism,
we changed the tidal criterion when the primary star is a
giant and during mass loss to be the same as it was when the primary star was on
the MS, $\rtide = 3 \Rinprim$. Under this new tidal criterion, the analogue of Figure \ref{fig:FigParAfter} is
presented in Figure \ref{fig:FigParAfternoRRG}. As expected, we see a sharp
decrease, from \TidalEvolvePer{} to \TidalEvolveRRGnonePer{}, in the fraction of triple systems that are tidally affected or collide when the
primary is a giant.  There are then many additional systems available after mass loss to increase the fraction of systems
that are tidally affected or collide after the primary becomes a WD.  The total change in the 
fraction is from \TidalAfterPer{} to \TidalAfterRRGnonePer{}.  Similarly, there is a significant increase in the fraction of 
MIEK systems, systems that collide or become tidally affected and flip the sign of \cosi, from \TidalFlipFirstAfterPer{} to \TidalFlipFirstAfterRRGnonePer{}.   For the example system considered, and for a uniform distribution
in eccentricities and mutual inclination, we consider \TidalFlipFirstAfterPer{} and
 \TidalFlipFirstAfterRRGnonePer{} to be lower and upper limits, respectively, to the fraction of all 
 systems that undergo MIEK.   

\section{Discussion}
\label{sec:discussion}

It is interesting to speculate on the outcome of triple systems that become tidally
affected at each of the various points in their evolution. In what follows, we discuss the possible outcomes for tidally interacting triple systems during the MS, giant, and WD 
evolutionary phases of the primary in \S \ref{sec:maindiscussion}, \S \ref{sec:giantdiscussion}, and \S \ref{sec:MIEKdiscussion}, respectively.  We then discuss the MIEK mechanism for massive star triples during NS formation of the primary in \S \ref{sec:ns}. Lastly, we discuss extensions of the current study in \S \ref{sec:FurtherStudies}.

\subsection{Main Sequence Phase}
\label{sec:maindiscussion}

We have shown in our fiducial parameter search, that systems that interact tidally 
while the primary is on the MS undergo both normal Kozai-Lidov oscillations and the 
eccentric Kozai mechanism in approximately equal proportion.  Those systems that 
undergo normal Kozai-Lidov oscillations are analogous to the systems considered by 
\citet{fabrycky07}, who calculated 
the long-term evolution of a distribution of triple systems using the 
quadrupole-order secular equations of  \citet{eggleton01}.  
\citet{fabrycky07} found that the combination of normal Kozai-Lidov cycles and tidal friction produces a substantial
population of close binaries, and that their calculations matched the observations of  \citet{tokovinin06}, which show 
that essentially all close spectroscopic binaries are actually triple systems.   Given these results, we anticipate that our systems 
with $\cos i_0\sim0$ will also tidally circularize with small \ain{} and \pin{}, with the caveat that the MS lifetime of 
the primary we consider here is much shorter than for the systems  considered by \citet{fabrycky07}, and
thus a fraction of our systems may not circularize before WD formation.   

Importantly, though, we find that half of the systems that tidally interact or collide on the MS do so as a result of the 
eccentric Kozai mechanism, which, as \citet{lithwick11}, \citet{katz11}, and \citet{naoz11a, naoz11b} have
shown cannot be captured by the quadrupole-order calculations of \citet{fabrycky07}.  What is the fate 
of these systems?  Unfortunately, no study has yet been performed to investigate the 
eccentric Kozai mechanism with the inclusion of tidal dissipation for triple stellar 
systems.  However, \citet{naoz11a} demonstrate with an octupole-order calculation that 
the eccentric Kozai mechanism and tidal friction can possibly explain the occurrence 
of retrograde hot Jupiters.  They show that the extremely high eccentricities obtained 
by the inner binary when \cosi{} flips signs can lead to a rapid capture of a planet 
into a short period retrograde orbit (``Kozai capture.'').  We suppose a similar 
mechanism operates for stellar triple systems, and that those that are tidally affected 
on the MS due to the eccentric Kozai mechanism rapidly dissipate energy from the inner 
binary and circularize with a semi-major axis of $a_1 \approx 2 \rperi\sim 2\rtide$.    
If so, then the \citet{fabrycky07} study may be missing
of order half of the systems that become close binaries.  

The action of the eccentric Kozai mechanism for MS binaries 
may also affect the distribution of binary eccentricities as a function of inner period as the
population is circularized \citep{socrates11, dong12}.  

An example observed 
(probable) intermediate mass triple star system that has likely already tidally circularized is 
MT429 in the Cygnus OB2 association, where the inner $P_1\simeq3$\,d B3V+B6V binary is orbited by 
a B0V tertiary at 138\,AU  \citep{kiminki12}.  An example of an even more massive 
system that might give rise to a NS-MS triple system as discussion in Section \ref{sec:ns} is HD 150136. 
It consists of O3-3.5V+O5.5-6V ($\sim64+40$\,M$_\odot$) inner binary with 
$P_1\simeq2.7$\,d and an O6.5-7V ($\sim35$\,M$_\odot$) tertiary with $P_2\sim3000-5000$\,d
\citep{mahy12}.

Finally,  after the inner binary orbit has shrunk due to either normal Kozai-Lidov 
oscillations or the eccentric Kozai mechanism, the binary evolves like other close
 binaries, potentially leading to the progenitors of single degenerate (SD) supernovae 
\citep{whelan73}, double degenerate (DD) supernovae \citep{webbink84, iben84}, barium 
stars (BA; \citealp{mcclure80}), and asymmetric planetary nebulae (PN) \citep{morris87}.

\subsection{Giant Phase}
\label{sec:giantdiscussion}

We find that $\sim30$\,\% of the triple systems we consider interact tidally or collide as the primary
evolves off the MS, but before mass loss, mostly as a result of normal Kozai-Lidov oscillations, which 
bring the pericenter of the secondary orbit inside $\sim$\,AU (see right panel, 
Fig.~\ref{fig:FigPar} and Equation \ref{eq:kozairperi3Giant}).  
Since the post-MS phase is short (particularly for the intermediate
and high mass stars we consider here), and since the efficiency
of tidal friction is uncertain, it is unclear if the system will circularize
before mass loss, or whether the Kozai-Lidov oscillation will simply be 
de-tuned such that pericenter $\sim$\,AU for the duration of the giant phase.  
The interaction of Kozai-Lidov oscillations as the primary evolves to 
the giant phase may also initiate mass transfer and complex 
secular dynamics, perhaps analogous to the system considered by 
\citet{prodan12}.   Another possibility is that this phase 
initiates common envelope evolution, perhaps circularizing
the inner orbit at smaller semi-major axis.
If the inner binary maintains significant eccentricity, or if
\ein{} is pumped after the giant phase and during mass loss 
via normal  Kozai-Lidov oscillations or MIEK, such systems
may form off-center or bi-polar planetary nebulae (PNe)
(e.g., \citealp{soker98, soker00, witt09}).
On the other hand, if the inner binary is circularized, it might also
participate in the shaping of PNe (e.g., \citealp{mastrodemos98, mastrodemos99}).  An example of some of 
these mechanisms in action may be ring planetary nebula SuWt 2 (Exter et al.~2010).
Since a very large fraction of triple
stars will be first affected by Kozai-Lidov cycles when the MS star 
evolves, this phase should be investigated in detail.

\begin{figure}[t]
	\centerline{
		\includegraphics[width=6.5cm, angle=90]{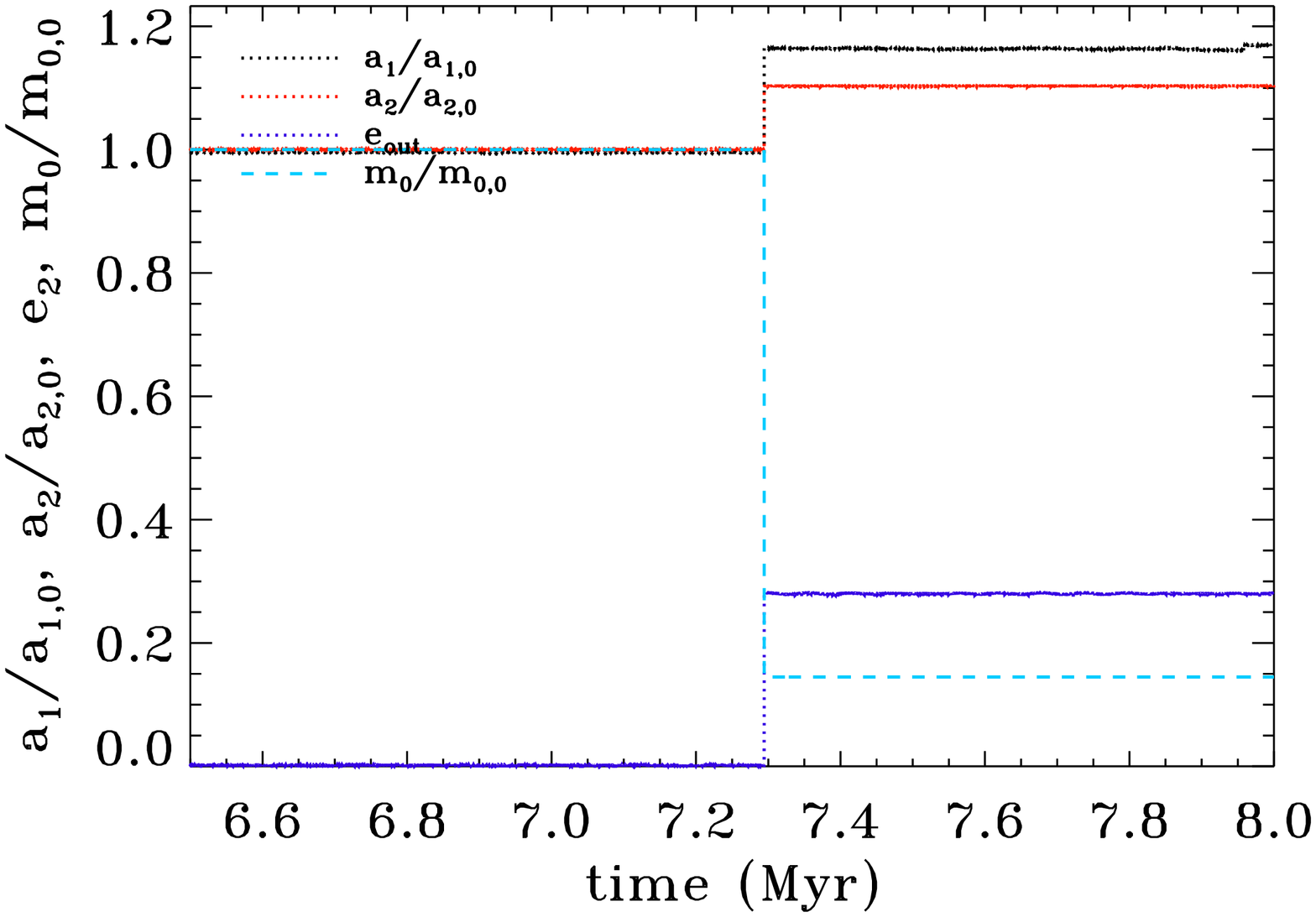}
	}
	\centerline{
		\includegraphics[width=6.5cm, angle=90]{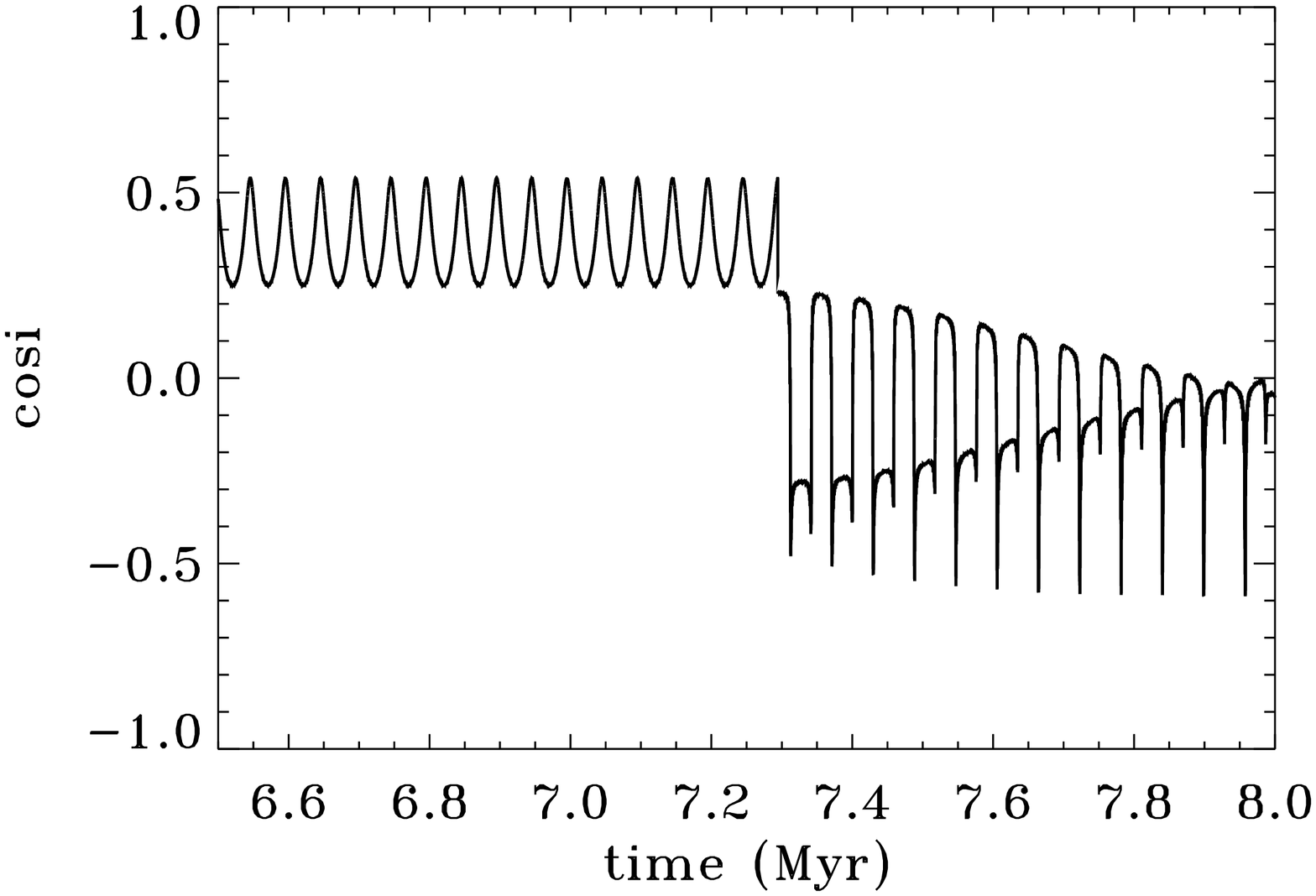}
	}
	\centerline{
		\includegraphics[width=6.5cm, angle=90]{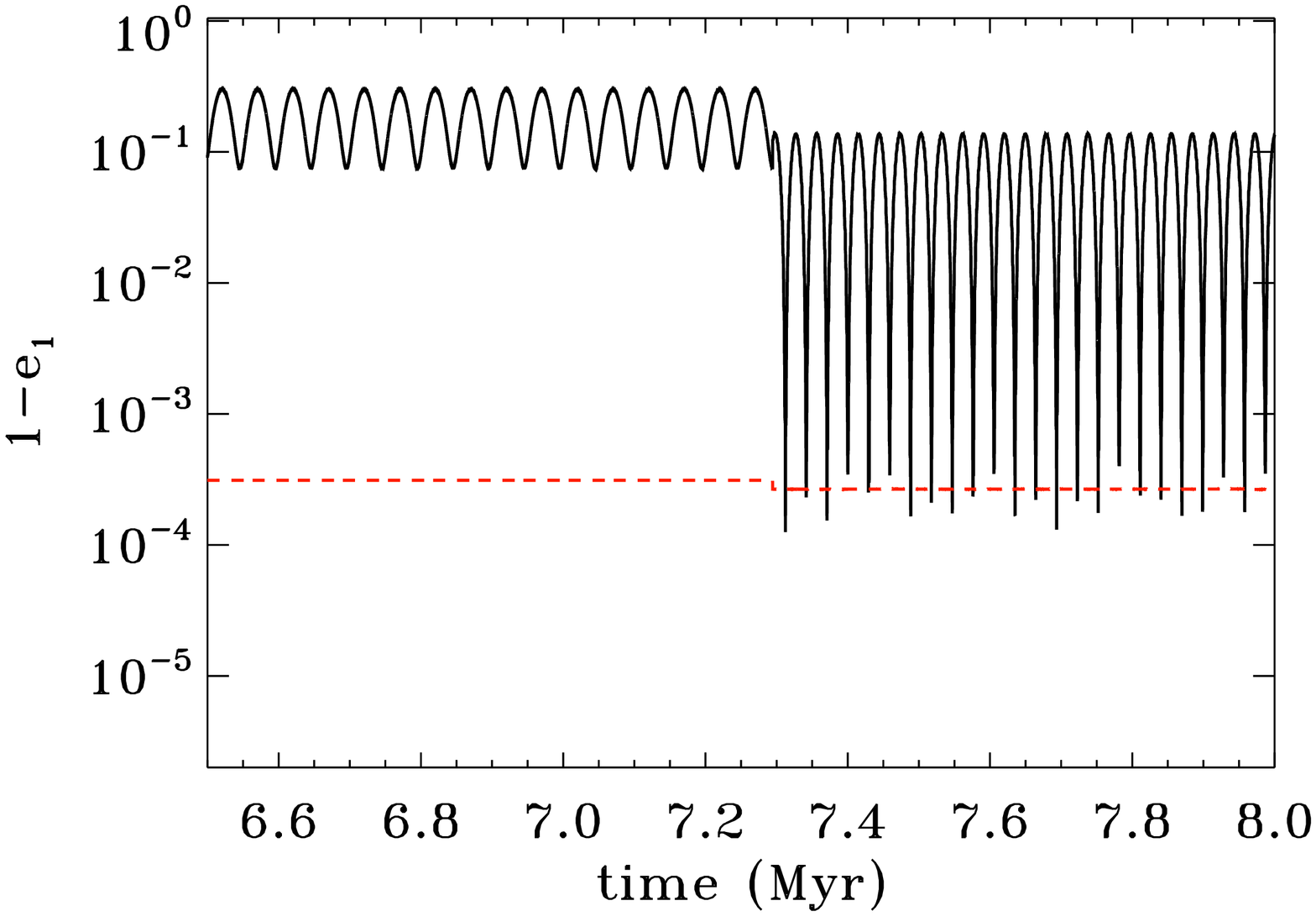}
	}
	\caption{MIEK for massive stars, similar to
Figure \ref{fig:FigExampleSystems}, but here $m_0=11$\,M$_\odot$,  
$m_1=10$\,M$_\odot$,  $m_2=10$\,M$_\odot$, $e_1=0.7$, $e_2=0$, $\cos i=0.25$,
$g_1=0\degree$, $g_2=270\degree$, $a_1=15$\,AU, and $a_2=300$\,AU. At $\sim7$ Myr, 
$m_0$ decreases to 1.4\,M$_\odot$, $\cos i$ flips, and the system attains
$1-e_1\sim10^{-3}$ immediately.}
	\label{fig:ns}
\end{figure}

\subsection{After Mass Loss \& MIEK}
\label{sec:MIEKdiscussion}

We find that $\sim2$\,\% of our triple 
systems undergo the MIEK mechanism after WD formation
(see Fig.~\ref{fig:FigExampleSystems}).
 We again emphasize that by removing all systems
from our sample that were tidally affected during the giant phase, the fraction of triples
that go through MIEK  presented in Table \ref{tab:ParmsOutcomes} (``After Mass-Loss")
is conservative since many of these systems may be available to go through MIEK.  The
reason for this is that as the primary enters its post-MS evolution and 
begins to interact tidally with the secondary, the orbits  will be affected
by processes that we do not model (e.g., \citealp{prodan12}).
 We attempt to provide an upper limit on the fraction of
triple systems that could be tidally affected or collide after mass-loss due to the MIEK
mechanism in Section \ref{sec:rtide}  by assuming a small
value for $r_{\rm tide}$ during the giant phase
(compare Figs.~\ref{fig:FigParAfter} and \ref{fig:FigParAfternoRRG}).  This
increases the fraction of triple systems that undergo
MIEK from \TidalFlipFirstAfterPer{} to \TidalFlipFirstAfterRRGnonePer{}.
A calculation of the  3-body dynamics with tides appropriate to a MS
star interacting with a red giant is clearly needed to address this issue 
in more detail and to address the issues raised in Section \ref{sec:giantdiscussion}.
  
Systems that undergo the MIEK mechanism also have many possible outcomes. 
One possibility is that the extreme eccentricities obtained during a ``flip'' might lead 
to physical collisions (Baoz et al. in prep.).  This may occur if, in a single orbit of the inner binary, \rperi{} changes from $\rperi > \rtide$ to $\rperi < R_{*}$. This is possible because the angular momentum of the inner binary at such large \ein{} is small.  Thus, the change in angular momentum required for $\rperi{} \sim 0$ is also small.  Another possible outcome for the system is a tidal Kozai capture
as presented in \citet{naoz11a}.  The inner semi-major axis \ain{} will strongly decrease
because of tides, leaving a close MS-WD binary. 
When the secondary subsequently  evolves off the MS, the WD will either 
accrete or go through common envelope evolution.
Both of these scenarios could lead to single-
or double-degenerate (SD or DD) SNe Ia, cataclysmic variables, or other types of WD accretors
such as  AM CVn stars \citep{warner95},  which may produce faint ``.Ia'' supernovae 
\citep{bildsten07}.

One interesting consequence of this evolutionary scheme for producing close WD-MS
binaries is that it skips the initial common
envelope evolutionary phase usually required in binary evolution to produce a close WD-MS pair.  
In the MIEK mechanism the close binary is produced without common envelope,
and thus the WD produced should have a mass 
appropriate to a single star with the mass of the primary (e.g., \citealp{kalirai08}).  
 This stands in  contrast to the conventional picture
for the production of WD-MS and WD-WD binaries relevant for SD and DD 
Ia progenitors, where the WD growth is truncated by common envelope interaction with the 
close secondary and the mass of the resulting WD is smaller than one
would naively estimate from the single-star initial-final mass relation \citep{iben84}.
For SD SN Ia models, this means that less mass needs to be
accreted onto the WD from the secondary, leading to a shorter delay time between 
production of the system and explosion. 
For DD SN models, if the model presented in \citet{pakmor12} is accurate,
then the luminosity of the SN resulting from the merger of a WD-WD
binary is primarily determined by the mass of the primary.  Thus, all else
being equal, if a WD-WD binary results from a MIEK triple system then its 
primary WD mass should be larger and thus result in a more luminous DD SN than the analogous close binary system
that is produced through normal common envelope evolution.

\subsection{Massive Star Triples \& Neutron Star Formation}
\label{sec:ns}

The MIEK mechanism may also play an important role in triple systems after NS
formation, particularly since the NS mass, $\simeq1.4$\,M$_\odot$, is so much less
than its progenitor massive star, leading to a large increase in $\epsilon_{\rm oct}$
at the time of the associated supernova (Equation \ref{eq:epsoct}).  However, several
effects differ in the NS case versus the case of WD formation considered throughout
this paper.  First, NSs receive a ``kick'' 
at birth that may unbind the inner binary or the outer tertiary, depending
on the masses of the constituents, their eccentricities, and their semi-major axes
(\citealp{hills83}; \citealp{kalogera96}).  Second, mass loss on the MS can affect the mass ratios as a function of time (e.g., \citealp{ekstrom11}).
Third, even in the absence of a NS kick, the large and instantaneous
mass loss in the supernova explosion from the primary provides a kick to the center of mass of the inner binary which may unbind the 
tertiary, and shift the mutual inclination.  Even so, we expect a population 
of NS-MS binaries to survive with massive tertiaries and large \eout{} due to the aforementioned kicks.  Thus, \epsoct{} will be large for the surviving triple systems which will cause many to exhibit the MIEK mechanism.  This is a qualitatively new way to produce close NS-MS binaries.

As a test, we integrated the orbit of a system with $m_0=11$\,M$_\odot$,  
$m_1=10$\,M$_\odot$,  $m_2=10$\,M$_\odot$, $e_1=0.7$, $e_2=0$, $\cos i=0.25$,
$g_1=0\degree$, $g_2=270\degree$, $a_1=15$\,AU, and $a_2=300$\,AU.  
At a time $\sim 7$ Myr after the start of the simulation we instantaneously decrease the 
mass of the primary to $1.4$\,M$_\odot$ and apply no kick to the 
NS.  Figure \ref{fig:ns} shows the resulting evolution.  The system 
flips immediately after the supernova and evolves to an inner eccentricity of $1-e_1 > 10^{-3}$, bringing the 
NS to tidal contact with the secondary.  Note the large increase in $e_2$
from $\sim0$ to $\sim0.3$ immediately after the SN. This calculation is meant 
only to be illustrative since it does not include a NS kick.  A full population study, with kicks and tides is
clearly needed to assess the viability of this mechanism for forming
close NS-MS binaries.

A further aspect of this type of evolution is that the ratio of 
$\aout / \ain$ can decrease during the strong MS mass loss and the 
supernova explosion,  which can cause the triple to become dynamically
unstable and eventually disrupt, or potentially initiate collisions
(the TEDI mechanism of \citealp{perets12}).

\subsection{Further Studies}
\label{sec:FurtherStudies}

In this study we have explored a single set of \minprim{} and \minsec{}, however, different combinations of masses will affect the fraction of systems which become tidally affected or collide during each of the evolutionary phases.  For example, if $\minprim \approx \minsec$ the system will have little or no time after the primary becomes a compact object before the secondary enters its giant phase.  This will reduce the fraction of systems that will flip and become tidally affected or collide due to the MIEK mechanism.  However, if the difference between the masses of the inner binary members is significantly greater, then a larger fraction of systems will flip before the primary goes through mass-loss.  Additionally, the difference in the masses after the primary has become a compact object will be smaller, further decreasing the fraction of systems that will be affected by the MIEK mechanism.  Determining the optimal balance between these two effects is beyond the scope of the current study. 

In a future study we will present a more general study of mass loss in a full population of
triples, quantifying the number of systems affected by Kozai-Lidov oscillations
and the eccentric Kozai mechanism, an accounting of those that are unbound and those
that are tidally affected or collide at each evolutionary stage, and those that undergo the MIEK mechanism.

Any study which evolves a full population of triple systems with distributions in \ain{} and \aout{} will have a significant fraction of systems that will be affected by the TEDI.  \citet{perets12} demonstrate that systems that start with a small $\aout / \ain$ ratio become unstable during mass-loss and evolve chaotically, leading to close encounters, collisions, and exchanges between the stellar components.  Additionally, \citet{perets12} also raise the issue of Kozai cycles and stellar evolution as important. Future studies will be needed to quantitatively evaluate the importance of both the TEDI and the MIEK mechanism.

\section{Conclusion}
\label{sec:conclusion}

We have presented a preliminary investigation of stellar evolution and mass loss in triple systems 
and demonstrated the MIEK mechanism, a novel channel through which triple systems produce
compact object -- MS binaries that interact tidally or collide only after the primary
has evolved off the MS and become a compact object.  

For a broad range of parameters, hierarchical triple star systems
are unaffected by Kozai-Lidov oscillations until the primary in the
central binary evolves off the MS and begins mass loss. Subsequently,
the primary becomes a WD or a NS, and may then be much
less massive than the other components in the ternary, enabling the ``eccentric Kozai
mechanism''. In this near-test-particle limit, the mutual inclination between
the inner and outer binary can flip signs, driving the inner binary to very high
eccentricity and tidal contact or collision. 
In this study, we define tidal contact by an ad-hoc minimal separation between the member of the inner binary, below which tidal affects are deemed important
 (Section \ref{section:tidal_criterion}).
 Even distant binaries with initial semi-major
axes larger then tens of AU  can be strongly affected.  We
consider an example triple system with masses 7, 6.5 and 6 \msun{}
as proof -of-principle (see Fig.~\ref{fig:FigExampleSystems}), and explore the MIEK
mechanism's dependence on the initial eccentricities and
inclination for this system.  For a flat distribution of eccentricities and $\cos i$, we find
that \TidalBeforePer{}, \TidalEvolvePer{}, and \TidalAfterPer{} of systems are tidally
affected or collide before the primary evolves off the MS, after the primary
becomes a giant, and after the primary becomes a WD, respectively. 

On the MS, we find that roughly half of our systems interact
tidally or collide as a result of the eccentric Kozai mechanism, and hence
their dynamics would not have been captured by quadrupole-order
secular calculations.  In the giant phase, most of the systems 
are brought to tidal contact by normal Kozai-Lidov oscillations 
and the fact that pericenter passage occurs at semi-major axes 
less than $\sim$\,AU even for very modest inclinations.  The 
large fraction of such systems motivates a detailed study of 
the dynamics and tidal interaction of triple systems with a giant 
primary.  Finally, the MIEK mechanism dominates the last stage, causing the WD produced by the 
initial primary to come to tidal contact or collide with the initial secondary
which reduces \ain{} without a common envelope phase. As a last application, we showed that some massive star triple systems will also undergo the MIEK mechanism after the primary's supernova explosion and the associated rapid mass loss (Figure \ref{fig:ns}).

We save a detailed study of the parameter space of triples affected by MIEK for a future paper,
but here note that for a thermal distribution of eccentricities, as is commonly 
adopted for studies of binary stars, we expect more 
systems to be affected by MIEK since, in general, $\epsilon_{\rm oct}$ increases
as $e_2$ increases (Equation \ref{eq:epsoct}).

\acknowledgments

We thank Chris Kochanek, Boaz Katz, Jennifer van Saders, Joe Antognini, Krzysztof Stanek,
Smadar Naoz, Yoram Lithwick, Norm Murray, and  Andy Gould for discussions and encouragement.  We also thank J. Fregeau for discussions and for making the code FEWBODY publicly available. This work is supported in part by an Alfred P. Sloan Foundation Fellowship and NSF grant AST-0908816.  B.J.S. was supported by a Graduate Research Fellowship from the National Science Foundation.

\bibliographystyle{apj}

\end{document}

%% file: Parms.tex
\newcommand{\ProTidalBefore}{$0.039$}
\newcommand{\ProTidalEvolve}{$0.30$}
\newcommand{\ProTidalDuring}{$0.00013$}
\newcommand{\ProTidalAfter}{$0.030$}

\newcommand{\ProTidalBeforePer}{$3.9\% $}

\newcommand{\ProTidalAfterPer}{$3.0\% $}
\newcommand{\ProTidalAfterPriorPer}{$33\% $}

\newcommand{\ProFlipBefore}{$0.0076$}
\newcommand{\ProFlipEvolve}{$0.000063$}
\newcommand{\ProFlipDuring}{$0.0020$}
\newcommand{\ProFlipAfter}{$0.029$}

\newcommand{\ProTidalFlipBefore}{$0.0059$}
\newcommand{\ProTidalFlipEvolve}{$0.000042$}
\newcommand{\ProTidalFlipDuring}{$0.000021$}
\newcommand{\ProTidalFlipAfter}{$0.028$}

\newcommand{\ProDistruptedBefore}{$0.15$}
\newcommand{\ProDistruptedEvolve}{$0.00042$}
\newcommand{\ProDistruptedDuring}{$0.0037$}
\newcommand{\ProDistruptedAfter}{$0.0083$}

\newcommand{\RetTidalBefore}{$0.22$}
\newcommand{\RetTidalEvolve}{$0.26$}
\newcommand{\RetTidalDuring}{$0.00020$}
\newcommand{\RetTidalAfter}{$0.015$}

\newcommand{\RetTidalBeforePer}{$22\% $}

\newcommand{\RetTidalAfterPer}{$1.5\% $}
\newcommand{\RetTidalAfterPriorPer}{$49\% $}

\newcommand{\RetFlipBefore}{$0.12$}
\newcommand{\RetFlipEvolve}{$0.000020$}
\newcommand{\RetFlipDuring}{$0.00097$}
\newcommand{\RetFlipAfter}{$0.013$}

\newcommand{\RetTidalFlipBefore}{$0.11$}
\newcommand{\RetTidalFlipEvolve}{$0.000020$}
\newcommand{\RetTidalFlipDuring}{$0.000040$}
\newcommand{\RetTidalFlipAfter}{$0.010$}

\newcommand{\RetDistruptedBefore}{$0.082$}
\newcommand{\RetDistruptedEvolve}{$0.00032$}
\newcommand{\RetDistruptedDuring}{$0.0017$}
\newcommand{\RetDistruptedAfter}{$0.014$}

\newcommand{\TidalBefore}{$0.13$}
\newcommand{\TidalEvolve}{$0.28$}
\newcommand{\TidalDuring}{$0.00016$}
\newcommand{\TidalAfter}{$0.022$}

\newcommand{\TidalBeforePer}{$13\% $}
\newcommand{\TidalEvolvePer}{$28\% $}
\newcommand{\TidalDuringPer}{$0.016\% $}
\newcommand{\TidalAfterPer}{$2.2\% $}

\newcommand{\FlipBefore}{$0.064$}
\newcommand{\FlipEvolve}{$0.000041$}
\newcommand{\FlipDuring}{$0.0015$}
\newcommand{\FlipAfter}{$0.021$}

\newcommand{\TidalFlipBefore}{$0.060$}
\newcommand{\TidalFlipEvolve}{$0.000030$}
\newcommand{\TidalFlipDuring}{$0.000030$}
\newcommand{\TidalFlipAfter}{$0.019$}

\newcommand{\TidalFlipBeforePer}{$6.0\% $}

\newcommand{\TidalFlipAfterPer}{$1.9\% $}

\newcommand{\TidalFlipFirstAfterPer}{$1.9\% $}

\newcommand{\DistruptedBefore}{$0.11$}
\newcommand{\DistruptedBeforePer}{$11\% $}
\newcommand{\DistruptedEvolve}{$0.00037$}
\newcommand{\DistruptedEvolvePer}{$0.037\% $}
\newcommand{\DistruptedDuring}{$0.0027$}
\newcommand{\DistruptedDuringPer}{$0.27\% $}
\newcommand{\DistruptedAfter}{$0.011$}

%% file: ParmsRtide_2.5.tex
\newcommand{\TidalAfterRtideless}{$0.022$}

%% file: ParmsRtide_3.5.tex
\newcommand{\TidalAfterRtidemore}{$0.023$}

%% file: ParmsRtide_5.tex
\newcommand{\TidalAfterRtidefive}{$0.023$}

%% file: ParmsRtide_10.tex
\newcommand{\TidalAfterRtideten}{$0.025$}

%% file: ParmsnoR_RG.tex
\newcommand{\TidalEvolveRRGnonePer}{$0.011\% $}

\newcommand{\TidalAfterRRGnonePer}{$8.7\% $}

\newcommand{\TidalFlipFirstAfterRRGnonePer}{$6.7\% $}

%% file: tab1.tex
\begin{deluxetable}{lllll}
\tablecolumns{5}
\tablewidth{0pc}
\tabletypesize{\scriptsize}
\tablecaption{Triple Outcomes}
\tablehead{
\colhead{} &
\colhead{Contact} &
\colhead{Flip} &
\colhead{Contact \& Flip} &
\colhead{Unbound} 
}
\startdata
{\bf Combined}  &              \\
$ \ \ \  \  $Primary MS & \TidalBefore{} & \FlipBefore{} & \TidalFlipBefore{} &  \DistruptedBefore{} \\
$ \ \ \  \  $Primary Giant & \TidalEvolve{} & \FlipEvolve{} & \TidalFlipEvolve{} & \DistruptedEvolve{} \\
$ \ \ \  \  $During Mass-Loss & \TidalDuring{} & \FlipDuring{} & \TidalFlipDuring{} & \DistruptedDuring{} \\
$ \ \ \ \   $After Mass-Loss & \TidalAfter{} & \FlipAfter{} & \TidalFlipAfter{} &  \DistruptedAfter{} \\
{\bf Prograde}  &              \\
$ \ \ \  \  $Primary MS & \ProTidalBefore{} & \ProFlipBefore{} & \ProTidalFlipBefore{} &  \ProDistruptedBefore{} \\
$ \ \ \  \  $Primary Giant & \ProTidalEvolve{} & \ProFlipEvolve{} & \ProTidalFlipEvolve{} & \ProDistruptedEvolve{} \\
$ \ \ \  \  $During Mass-Loss & \ProTidalDuring{} & \ProFlipDuring{} & \ProTidalFlipDuring{}  & \ProDistruptedDuring{} \\
$ \ \ \  \  $After Mass-Loss & \ProTidalAfter{} & \ProFlipAfter{} & \ProTidalFlipAfter{} & \ProDistruptedAfter{} \\
{\bf Retrograde}  &              \\
$ \ \ \  \  $Primary MS & \RetTidalBefore{} & \RetFlipBefore{} & \RetTidalFlipBefore{} &  \RetDistruptedBefore{} \\
$ \ \ \  \  $Primary Giant & \RetTidalEvolve{} & \RetFlipEvolve{} & \RetTidalFlipEvolve{} & \RetDistruptedEvolve{} \\
$ \ \ \  \  $During Mass-Loss & \RetTidalDuring{} & \RetFlipDuring{} & \RetTidalFlipDuring{} & \RetDistruptedDuring{} \\
$ \ \ \  \  $After Mass-Loss & \RetTidalAfter{} & \RetFlipAfter{} & \RetTidalFlipAfter{} & \RetDistruptedAfter{}

\enddata
\tablecomments{Fractions of triple systems that become tidally affected, exhibit
a flip, or are unbound at various evolutionary stages without being tidally affected or unbound in a previous evolutionary stage.  Prograde and
retrograde divide the sample in half where combined are the fractions of the
whole sample. The rows Primary MS, Primary Giant, During Mass-Loss and After
Mass-Loss present the fractions of triple systems with the primary star is on
the main sequence, a giant that has evolved off the main sequence, an evolved
star undergoing an $10^4$ year phase of significant mass-loss phase and a WD,
respectively. The columns are as follows.  {\it Contact ---} Systems that satisfy our ad-hoc tidal criteria.  
{\it Flip ---} Systems whose \cosi{} changes sign from its initial value. 
{\it Contact \& Flip ---} Systems common to the previous two columns. 
{\it Unbound ---} Systems that become unbound without being tidally affected prior.  
Table discussed in \S\ref{sec:parameters}. 
}

\label{tab:ParmsOutcomes}
\end{deluxetable}